\documentclass[aps,prd,twocolumn,superscriptaddress,showpacs]{revtex4}
\usepackage{graphicx}
\usepackage{dcolumn}
\usepackage{bm}
\usepackage{natbib}

\newcommand{\be}{\begin{equation}}
\newcommand{\ee}{\end{equation}}
\newcommand{\bea}{\begin{eqnarray}}
\newcommand{\eea}{\end{eqnarray}}

\newcommand{\sv}{\langle\sigma_Av\rangle}

\newcommand{\eV}{{\rm ~eV}}

\newcommand{\keV}{{\rm ~keV}}

\newcommand{\GeV}{{\rm ~GeV}}

\newcommand{\sigmav}{\langle\sigma_Av\rangle}

\def\ga{\vcenter{\hbox{$>$}\offinterlineskip\hbox{$\sim$}}}

\begin{document}
%\draft

\title{CMB Constraints on WIMP Annihilation: Energy Absorption During the Recombination Epoch}

\author{Tracy R. Slatyer}
\email{tslatyer@fas.harvard.edu}
\affiliation{Physics Department, Harvard University, Cambridge, MA 02138, USA}

\author{Nikhil Padmanabhan}
\email{NPadmanabhan@lbl.gov}
\affiliation{Physics Division, Lawrence Berkeley National Laboratory, 1
  Cyclotron Rd., Berkeley, CA 94720, USA}

\author{Douglas P. Finkbeiner$^{1,}$}
\email{dfinkbeiner@cfa.harvard.edu}
\affiliation{Harvard-Smithsonian Center for Astrophysics, 60 Garden St., Cambridge, MA 02138, USA}

%\date{\today}

\begin{abstract}We compute in detail the rate at which energy injected by dark matter annihilation heats and ionizes the photon-baryon plasma at $z \sim 1000$, and provide accurate fitting functions over the relevant redshift range for a broad array of annihilation channels and DM masses.
The resulting perturbations to the ionization history can be constrained by measurements of the CMB temperature and polarization angular power spectra.
We show that models which fit recently measured excesses in 10-1000 GeV electron and positron cosmic rays are already close to the $95 \%$ confidence limits from WMAP. The recently launched Planck satellite will be capable of ruling out a wide range of DM explanations for these excesses. In models of dark matter with Sommerfeld-enhanced annihilation, where $\langle \sigma v \rangle$ rises with decreasing WIMP velocity until some saturation point, the WMAP5 constraints imply that the enhancement must be close to saturation in the neighborhood of the Earth. 
\end{abstract}


\maketitle

\section{Introduction}

Dark matter (DM) annihilation around the redshift of last scattering ($z \sim 1000$) can modify the observed temperature and polarization fluctuations of the CMB, which have been measured to high precision by experiments such as WMAP, ACBAR and BOOMERANG \cite{Komatsu:2008hk, Reichardt:2008ay,Montroy:2005yx}. Other indirect astrophysical probes of DM annihilation must contend with the complexities of Galactic astrophysics -- for example, the DM distribution
and clumpiness, ISM density, magnetic field strength
and degree of tangling, Galactic photon energy density,
etc. All of these are complex processes with significant
uncertainties; in contrast, the mechanisms by which DM annihilation modifies the CMB are relatively simple and well understood.

DM annihilation injects high-energy particles into the IGM \footnote{The use of
``Inter-Galactic Medium'' for the photon-baryon fluid at $z\sim 1000$ is
a convenient anachronism; we use it throughout without further apology.}, which heat and ionize neutral hydrogen as they cool. This ionizing energy does not generally change the redshift of recombination, but does alter the residual ionization after recombination. The increased ionization fraction leads to a broadening of the last scattering surface, attenuating correlations between temperature fluctuations. The low-$\ell$ correlations between polarization fluctuations, on the other hand, are enhanced by the thicker scattering surface.
 
These effects of WIMP  
annihilations on recombination have been studied previously, with  
significant effects on WMAP for $\sigmav \sim (10^{-26}/f) \times (M_\mathrm{DM}/(1 \mathrm{GeV}/c^2))$ cm$^3$/s and on
Planck for $\sigmav \sim (10^{-27}/f) \times (M_\mathrm{DM}/(1 \mathrm{GeV}/c^2))$ cm$^3$/s, where $f$ is an unknown parameter of order $\sim 0.01-1$ \cite{Padmanabhan:2005es}. For $\mathcal{O}$(TeV) WIMPs, these cross sections were considerably larger than the standard thermal cross  sections expected at the time, but in light of recent  
cosmic-ray experiments which motivate larger annihilation cross sections, it is important to reconsider these limits  
and quantify the free parameter $f$. 

The PAMELA, ATIC, PPB-BETS, Fermi and H.E.S.S experiments have observed unexpected features in the electron and positron cosmic-ray spectra, at energies of 10-1000 GeV. PAMELA \cite{Adriani:2008zr} has measured the positron flux ratio $\phi(e^+)/(\phi(e^+)+\phi(e^-))$ and found a sharp rise starting at 10 GeV and continuing up to $\sim 100$ GeV. This confirms previous indications of
an excess by HEAT \cite{Barwick:1997ig} and AMS-01 \cite{ams}, but at
much higher confidence. The ATIC balloon experiment \cite{ATIClatest} has measured the spectrum of $e^++e^-$ (ATIC
cannot distinguish positrons from electrons) from 20-2000 GeV, and finds a broad excess at $300-800$ GeV, in agreement with the similar excess
observed by PPB-BETS
\cite{Torii:2008xu}. The Fermi \cite{Abdo:2009zk} and H.E.S.S \cite{Aharonian:2009ah} experiments have measured a similar but somewhat smaller $e^+ + e^-$ excess in the $300-1000$ GeV energy range, and have not confirmed the peak and sharp cutoff observed by ATIC around 700 GeV. However, in combination, these results argue for a new primary source of high-energy electrons and positrons. 

The annihilation of WIMPs in the Galactic halo has been proposed as this new source, with the other major contender being $e^+ e^-$ pairs from pulsars \cite{pulsars,2001A&A...368.1063Z,Hooper:2008kg,Yuksel:2008rf}. While weak-scale dark matter annihilating in the Galactic halo can generically produce annihilation products with energies in the hundreds of GeV, including electrons and positrons, most DM annihilation channels give rise to insufficiently hard electron spectra. For energies up to 100 GeV, PAMELA observed no antiproton excess \cite{Adriani:2008zq}, strongly constraining annihilation channels that produce copious antiprotons. Gamma ray bounds also constrain the production of neutral pions (either directly or as products of a hadronic cascade). Taking these constraints into account, for dark matter masses in the 10 GeV - 1 TeV range, the WIMP must annihilate primarily to leptons in order to fit PAMELA observations \cite{Cholis:2008hb, Cirelli:2008pk, Cholis:2008qq}.  

Even for these primarily leptonic annihilation channels, the annihilation cross section required to fit the observed excesses is significantly higher than the thermal relic cross section \cite{Cholis:2008qq, Cholis:2008wq}. For a WIMP with  $\sim$ TeV mass, as suggested by the measured charge-undifferentiated electron spectrum, a boost factor of 2-3 orders of magnitude is required, depending on the annihilation channel. Lower masses allow smaller boost factors, but a 100 GeV WIMP still requires a boost factor of $\sim 10$ or higher relative to the thermal relic cross section, to produce the PAMELA positron excess (and of course, cannot produce the observed ATIC/PPB-BETS/Fermi/H.E.S.S excesses).

Several authors have suggested models of thermal relic DM where $\sigmav$ rises with decreasing WIMP velocity as a consequence of a \emph{Sommerfeld enhancement}, and thus the present-day DM annihilation cross section is considerably higher than the thermal relic cross section $\sigmav \sim 3 \times 10^{-26}$ cm$^3$/s \cite{Cirelli:2007xd, MarchRussell:2008yu, ArkaniHamed:2008qn, Pospelov:2008jd, Nomura:2008ru}. Sommerfeld enhancement occurs when some particle with mass much smaller than the WIMP mass mediates a long-range attractive force between the WIMPs \cite{sommerfeld, Hisano:2004ds}. For heavy WIMPs ($\sim 10$ TeV), these force carriers can simply be the Standard Model $W$ and $Z$ bosons \cite{Cirelli:2007xd}. Other models invoke new light dark-sector particles \cite{Nomura:2008ru, ArkaniHamed:2008qn, Pospelov:2008jd} to mediate a Sommerfeld enhancement for weak-scale DM ($\sim 100 - 1000$ GeV).  In general, Sommerfeld-enhanced models suggest that a much higher annihilation cross section around the redshift of recombination may be possible.

Sommerfeld-enhanced dark matter models with weak-scale DM masses and GeV-scale light force carriers are especially attractive for explaining the observed cosmic ray anomalies \cite{ArkaniHamed:2008qn, Pospelov:2008jd}. If the DM annihilates to the light force carriers, then decays into protons/antiprotons (and gauge bosons, etc) are kinematically forbidden, and the annihilation products are predominantly hard leptons \cite{Cholis:2008vb}. The electron and positron spectra produced by such annihilation channels provide excellent fits to the ATIC and PAMELA $e^+ e^-$ spectra, up to a boost factor attributed to the Sommerfeld enhancement \cite{Cholis:2008wq}, and similarly fit the Fermi data well \cite{Meade:2009iu}. Such models have also been proposed to explain anomalous results from the DAMA \cite{Bernabei:2008yi} experiment with inelastic WIMP-nucleon scattering \cite{Smith:2001hy, Tucker-Smith:2004jv, Chang:2008gd}, and the
INTEGRAL \cite{Weidenspointner:2007} signal with inelastic WIMP-WIMP scattering \cite{Finkbeiner:2007kk, Pospelov:2007xh}, which naturally arises in this framework \cite{ArkaniHamed:2008qn}.
We adopt the terminology of \cite{Finkbeiner:2007kk}, and refer to these models as ``exciting dark matter'' (XDM). 
 CMB probes of DM annihilation are particularly well suited to constraining models with Sommerfeld-enhanced annihilation, since the $1/v$ scaling of $\sigmav$ results in a large annihilation cross section when the DM is diffuse and cold.

In order to constrain specific models of DM annihilation, it is essential to understand the efficiency with which energy from DM annihilation heats and ionizes the IGM, as it is this deposited energy which perturbs the ionization history and hence the CMB. Several previous studies have investigated the effect of dark matter annihilation/decay on the ionization history at recombination and during the cosmic ``dark ages'' \citep{Chen:2003gz, Pierpaoli:2003rz, Padmanabhan:2005es, Furlanetto:2006wp, Mapelli:2006ej, Zhang:2007zzh, Galli:2009zc}. However, these works have generally either assumed that the energy from annihilation/decay is deposited promptly with some redshift-independent efficiency, leaving this efficiency factor $f$ as a free parameter (degenerate with varying the annihilation cross section), or alternatively characterized the universe as opaque except for photons with energies lying in some ``transparency window''.

We calculate in detail the rate at which energy from dark matter annihilation is deposited into the IGM, via interaction of the annihilation products with the photon-baryon plasma. This allows us to apply previously derived constraints on $f \sigmav$ to the DM model itself; in particular, we can directly apply the WMAP5 constraints to models proposed to explain the observed cosmic-ray anomalies. 

A similar calculation has been performed by \cite{Ripamonti:2006gq}, and employed in \cite{Valdes:2007cu} to discuss the effect of light dark matter decays on the 21cm line. However, this analysis applied only to decay and annihilation of light ($\lesssim 10$ MeV) dark matter, where high energy photon cooling mechanisms such as pair production and photon-photon scattering could be neglected, and also assumed a near-monoenergetic spectrum. In contrast, our calculation is valid for arbitrary spectra of the DM annihilation products, for dark matter masses up to TeV scales. Our analysis also improves on the prior calculation by taking into account the changing ionization fraction of the universe around recombination, and the energy injection from DM annihilation products at redshifts greater than $z \sim 1100$, which cool slowly and eventually deposit their energy at a lower redshift.

Section \ref{sec:englosscode} details our numerical calculation of the energy deposition from DM annihilation, while Section \ref{sec:fcalc} presents our results for the redshift-dependent efficiency factor $f(z)$, for an array of annihilation channels. Section \ref{sec:cmbconstraints} discusses WMAP5 constraints on the annihilation cross section for various annihilation channels, the implications for DM models with Sommerfeld-enhanced annihilation, and the prospect that Planck \cite{Planck:2006uk} can rule out dark matter annihilation as an explanation for the observed cosmic-ray anomalies.

\section{Energy loss processes for DM annihilation products}
\label{sec:englosscode}

Depending on the DM model, the dark matter may annihilate to a wide range of particles:  gauge bosons, charged leptons, neutrinos, hadrons, or more exotic states. These annihilation products may subsequently decay or interact with the IGM, producing showers of $e^{\pm}$ pairs, protons and antiprotons, photons and neutrinos. Neutrinos are stable and weakly interacting, so they escape and their energy is lost, while protons are highly penetrating and poor at transferring energy to the IGM \cite{Chen:2003gz}. Heating and ionization of the IGM occurs primarily through the electrons, positrons, and photons injected as a result of DM annihilation, either directly or by subsequent decays and interactions of the primary annihilation products. Positrons behave identically to electrons at high energies, while at low energies they thermalize, form positronium and annihilate into photons. Thus the problem of computing the energy deposition efficiency reduces to calculating the evolution of the photon and electron spectra.

The specific annihilation channels we focus on in this paper are motivated by the anomalous excesses observed by PAMELA and ATIC, and involve annihilation either directly to charged lepton pairs, or to light dark-sector states which decay to charged lepton pairs \cite{Cholis:2008wq}. In the latter case, the same light dark-sector states mediate a Sommerfeld enhancement. We also compute the energy deposition efficiency for two benchmark masses (200 GeV and 1 TeV) for WIMPs annihilating via other SM channels.

\subsection{Electron cooling and energy deposition mechanisms}

Energy loss processes for high energy electrons have been considered by a number of authors
\cite{1970RvMP...42..237B, 1985ApJ...298..268S, Chen:2003gz}. The primary cooling mechanisms are inverse Compton scattering on CMB photons at high energy ($\gamma \gg 1$), and collisional heating, excitation and ionization at low energy (see Appendix \ref{app:electrons} for the detailed cross sections).

In all cases, electrons deposit their energy into the IGM, or produce high-energy photons via inverse Compton scattering on the CMB, on timescales short compared to the Hubble time. For high-energy electrons, most of the energy goes into inverse Compton scattered photons, converting the injected electron spectrum to an effective injected photon spectrum. It is important to take into account repeated inverse Compton scatterings, as only in the extreme Klein-Nishina limit do electrons lose all their energy in the first scattering, so high-energy electrons can produce many low-energy photons via repeated scatterings. Once these electrons have cooled to low energies where ICS is no longer the dominant process, they -- like the electrons injected at low energy -- rapidly deposit their remaining energy to the IGM by ionization, excitation and heating. The positrons produced by DM annihilation behave identically to electrons at high energies; at the lowest energies they form positronium and then annihilate, producing the usual positronium spectrum with a continuum up to 511 keV and a line at the cutoff.

In order to determine the fraction of annihilation power deposited to the IGM as a function of redshift, we must compute the amount of energy deposited to the IGM and the spectrum of photons produced by these processes, when the energy in the injected electron spectrum has been completely depleted. The relative rates of the various energy loss processes change with redshift, so we perform this calculation for an arbitrary injected electron spectrum at each of the relevant redshifts. As all these processes occur on timescales much faster than the Hubble time, we do not consider redshifting effects, and so the fraction of energy loss into each mode can be determined simply by comparing their rates. For electrons of each energy, we record the fraction of energy that goes into created or upscattered photons and energy deposition to the IGM, and the fraction retained by the electron as it downscatters to lower energies.

We assume that below a certain kinetic energy (presently set to $250$ eV) electrons efficiently deposit all their kinetic energy, cooling completely without producing photons by ICS. We also make the approximation that after a single ionization, an electron deposits all its energy, since in the energy range where ionization dominates inverse Compton scattering, electrons tend to lose a large fraction of their energy to ionization, and the resulting secondary electrons efficiently deposit their energy to the IGM \cite{1985ApJ...298..268S}).

For each initial electron energy, we can then determine the eventual partition of the electron's energy between produced photons and energy deposited to the IGM, when the electron has completely cooled and thermalized (and possibly annihilated). We use an inductive approach, starting with the lowest energy bins and working up in energy. Since we have recorded the fate of all electrons with lower initial energy, after the primary electron has downscattered once we can employ the previously calculated cooling histories.

Once the energy partitioning has been determined for all electron bins, integrating this result over the injected electron spectrum yields the prompt deposited energy and the photon spectrum produced by electrons injected at the given redshift. Determining the annihilation power that is eventually deposited to the IGM then becomes a question of how the photons deposit their energy. The dominant photon energy loss processes are not much faster than $H(z)$ over the entire relevant energy range, so redshifting must be taken into account. 

Applying the same techniques to electrons and positrons from pair production, or electrons from Compton scattering, allows us to view these mechanisms as converting a given photon spectrum to a new photon spectrum, plus some rapidly deposited energy. In this way we ``integrate out'' the fast electron cooling processes and reduce our numerical problem to computing the evolution of the photon spectrum only.

\subsection{Energy deposition from photons} 

\begin{figure}
\includegraphics[width=3in]{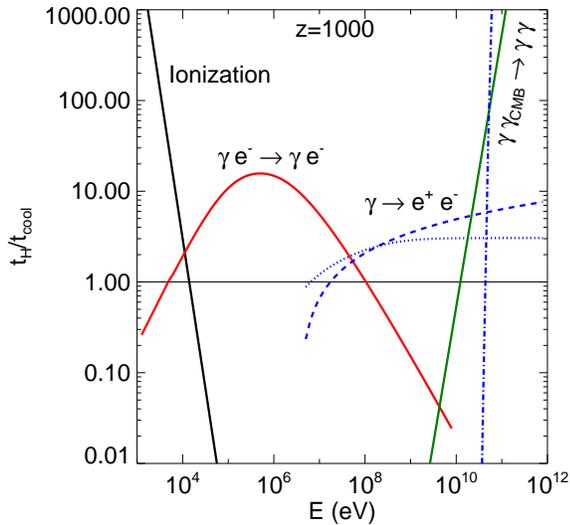}
\caption{\label{fig:photonlosses} A comparison of the photon cooling time to the Hubble time at $z=1000$,
for different photon energies. The dominant processes (in order of increasing energy) are ionization, 
Compton scattering, pair production on the H/He gas, photon-photon scattering, and pair production on the CMB. All the curves assume a He mass fraction of 1/4, with a  density of $2.57 \times 10^{-7}$ amu / cm$^{3}$ today. The dotted curve shows pair production on a neutral IGM, the dashed curve shows pair production on a fully ionized IGM, and the dashed-dotted curve represents pair production on the CMB.  This figure updates Fig. 1 in \cite{Padmanabhan:2005es}, which had an error leading to cooling times approximately a factor of three longer.}
\end{figure}

\begin{figure}
\includegraphics[width=3in]{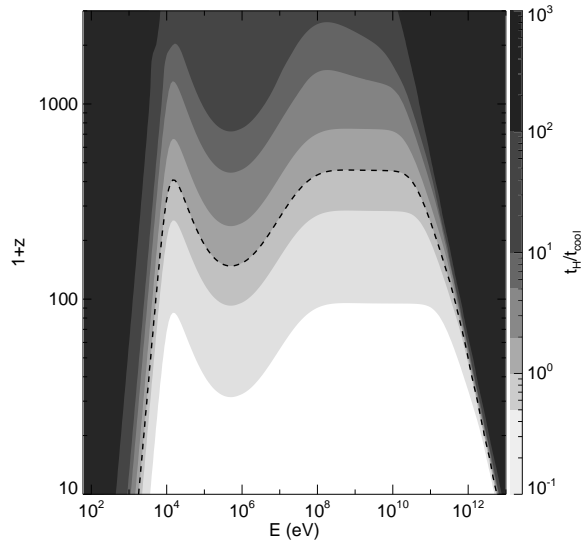}
\caption{\label{fig:transparencywindow} A comparison of the photon cooling time (from all processes) to the Hubble time over the entire redshift range of interest. The plot assumes a He mass fraction of 1/4, with a baryon density of $2.57 \times 10^{-7}$ amu / cm$^{3}$ today, and the standard ionization history and fiducial cosmology. The dashed line corresponds to $t_\mathrm{cool} = t_H$. There is a discrepancy between this figure and Fig. 2 in the originally published version of \cite{Chen:2003gz}: the authors of that paper have advised us that upon revising their calculation, their results now agree with ours.}
\end{figure}

The interaction of photons with the IGM was considered in detail by \cite{1989ApJ...344..551Z,1990ApJ...349..415S},
who find that the dominant processes (ordered by increasing photon energy) are photoionization, 
Compton scattering, pair production off nuclei and atoms, photon-photon scattering, and pair
production off CMB photons. The cross sections and spectra for these processes are listed in Appendix \ref{app:photons}. To estimate the efficiency of these mechanisms, we compare the
cooling time for each process, $t_\mathrm{cool} \equiv 1/(d \ln E/dt)$, to the Hubble time, $t_{H} \equiv 1/H(z)$.
Except for Compton scattering and photon-photon scattering, we approximate the cooling time by the mean free time as most of the energy is lost in the first interaction. If $t_{H} \gg t_\mathrm{cool}$, the photons ionize the IGM, produce energetic electrons, or downscatter, very rapidly. Conversely, if $t_{H} \ll t_\mathrm{cool}$, the universe is optically thin and most of the energy is lost 
through the redshifting of photons. The results of this comparison are shown in Fig. \ref{fig:photonlosses} and Fig. \ref{fig:transparencywindow}. At the relevant redshifts for hydrogen recombination, $z \sim 700-1200$, while the universe is not transparent at the relevant energies, it is also not sufficiently opaque that we can ignore redshift entirely.   

For photons with energies below $\sim 10^3$ eV and above $\sim 10^{11}$ eV, in the redshift range of interest, the dominant processes (photoionization and pair production on the CMB, respectively) take place on timescales much faster than the Hubble time. The lowest-energy photons deposit their energy into the IGM by photoionization, while the highest-energy photons rapidly pair produce or downscatter on the CMB. Photon-photon scattering is a ``photon splitting'' process that yields an approximately flat photon spectrum (up to the energy of the initial photon), whereas pair production produces an approximately flat spectrum of high-energy electrons and positrons which rapidly inverse Compton scatter to produce a softer photon spectrum. 

Photons lying in the broad $10^3-10^{11}$ eV range scatter or pair-produce on timescales within a few orders of magnitude of the Hubble time, while slowly redshifting away their energy. With decreasing redshift, all the energy loss processes decrease in efficiency relative to the Hubble time, as shown in Fig. \ref{fig:transparencywindow}. 

There is a ``transparency window'' at $\sim 10^{8} ~-~ 10^{10} \eV$ at $z=1000$, where the  cooling time of the dominant energy loss processes is close to the Hubble time. The ratio $t_{H}/t_\mathrm{cool} 
\propto (1+z)^{3/2}$ ($\propto (1+z)^{9/2}$ for photon-photon scattering), while the photon
energy redshifts as $(1+z)$: photons injected into a transparency window can therefore remain in the optically thin regime, 
and contribute to the diffuse photon background today. Below this energy range, Compton scattering rapidly depletes the photon spectrum, but becomes inefficient as an energy-loss process at lower energies where Compton scattering becomes purely elastic. The result is a second, narrower ''transparency window'' below the Compton bump.

\subsection{Beyond the ``on-the-spot'' approximation}
\label{subsec:onthespot}

Previous analyses of the effect of DM annihilation/decay on the ionization history of the universe \cite{Chen:2003gz, Padmanabhan:2005es, Furlanetto:2006wp, Galli:2009zc} have employed the ``on-the-spot'' approximation, where the energy from DM annihilation/decay is assumed to be instantaneously deposited in the IGM with some efficiency $f \sim 0.01-1$. The effects of Sommerfeld enhancement can be determined within these models simply by multiplying the DM annihilation cross section by the appropriate saturated enhancement factor.

However, the on-the-spot approximation is not necessarily well justified at redshift $\sim 1000$. The energy loss processes for electrons, very low-energy photons and very high-energy photons are all extremely rapid compared to the Hubble time, but over much of the relevant energy range the timescale for photon energy loss is of the same order as the Hubble time (Fig. \ref{fig:transparencywindow}). If a significant fraction of the energy from annihilation is injected at this energy scale or higher, assuming either that photons in the semi-transparent regime promptly deposit their energy or that they never deposit their energy may give a poor approximation to the actual deposition history.

Instead, we evolve the spectra of the photons from DM annihilation with the various energy loss processes described above, taking redshifting into account, and track the energy deposited as a function of redshift. However, at very high and low photon energies, the energy loss processes are very rapid compared to the Hubble time. Stable numerical evolution of the photon spectrum requires that we integrate out these rapid processes (or take extremely small timesteps, but this is not practical when the timescale for the fastest cooling mechanisms may be 12 orders of magnitude shorter than the Hubble time). 

At very low energies we may simply assume prompt deposition with $100 \%$ efficiency, but resolving the high-energy processes is more complicated. Although a 100 GeV photon will scatter or pair-produce very quickly relative to the Hubble time, this does \emph{not} imply that all its energy is deposited on that timescale: photon-photon scattering on the CMB will split the original photon's energy repeatedly until it enters the transparency window, and inverse Compton scattering on the electrons and positrons resulting from pair production can also inject high-energy photons into the transparency window. 

\subsection{Resolving rapid cooling processes for high energy photons}

To resolve the effect of high-energy photon cooling processes, we begin by ignoring redshifting entirely (since by construction, these processes occur on timescales several orders of magnitude shorter than a Hubble time). Our goal is to compute the total energy deposited to the IGM, and the resulting spectrum of lower-energy photons, once no photons remain at the highest energies, due to pair production and downscattering (which occurs quickly relative to a Hubble time).

We compare the rates of the various processes to determine the fraction of the initial photon's energy that is directly injected into the IGM, and the spectrum of lower-energy photons that is created, in the photon's first interaction. For high energy photons, the former occurs primarily by pair production followed by excitation, ionization or heating by the newly produced electrons or positrons. The latter occurs by (1) photon-photon scattering, and (2) pair production followed by inverse Compton scattering and positronium annihilation of the $e^+ e^-$ pairs. As described previously, we integrate out the electron cooling processes to determine the scattered photon spectrum and deposited energy resulting from pair production. 

Part of the resulting photon spectrum may still lie in the high energy region where the energy loss processes are too rapid (compared to $t_H$) to be easily included when calculating the evolution of the spectrum over cosmological timescales. Thus we need to apply the same procedure to the high energy bins in the new spectrum, determining the photon spectrum produced by downscattering and pair production from those bins. Iterating this process is equivalent to taking into account multiple interactions for the high-energy photons inside a single timestep, and the resulting infinite series can be written in terms of the sum of a geometric series of square matrices (of size equal to the number of high energy photon bins). The series can then be resummed analytically by a matrix inversion, which is computationally tractable so long as the number of high energy bins is not too large.

\subsection{Time evolution of the photon spectrum}

We employ a simple first-order integration scheme for the time evolution of the photon spectrum and energy deposited in the IGM. We initialize the photon spectrum as zero everywhere (we do not track the CMB photons), so there is an initial transient behavior due to the neglect of photons from earlier DM annihilations. However, at $z > 2000$ the deposition efficiency is excellent and the on-the-spot approximation is quite accurate, so the transient behavior dies away rapidly provided the initial redshift is sufficiently high. We choose an initial redshift of $z_\mathrm{init} = 4000$. We employ the standard ionization history computed by the publicly available code \texttt{RECFAST} \cite{Seager:1999}, and assume perturbations to the ionization history due to the extra injected energy to be small (since large perturbations to the ionization history would violate existing constraints).

At each timestep, we divide the photon spectrum into three regions: low energy photons where the (inelastic) interaction rate is short compared to $H(z)$ (``fast'' defined as a total rate of more than one interaction per photon per timestep, for timesteps typically of order $\sim t_H/1000$), high energy photons where the interaction rate is fast compared to $H(z)$, and an intermediate region where the interaction rates are no more than a few orders of magnitude greater than $H(z)$ and so the timestep is small enough to resolve them. We take a standard timestep of $d \ln (1 + z) = 10^{-3}$. This division is redshift-dependent and so must be performed at each timestep. The low energy photons are assumed to deposit all their energy within the timestep, and the high energy photon processes are integrated out as described above. The part of the photon spectrum in the intermediate energy range is evolved by the photon cooling mechanisms listed in Appendix \ref{app:photons}, and by redshifting. At the end of each timestep, the photon spectrum is updated and new photons are injected from DM annihilation (both from direct production, and from ICS and annihilation of electrons and positrons). The energy deposited to the IGM at each step is recorded.

\begin{figure}
\includegraphics[width=3.5in]{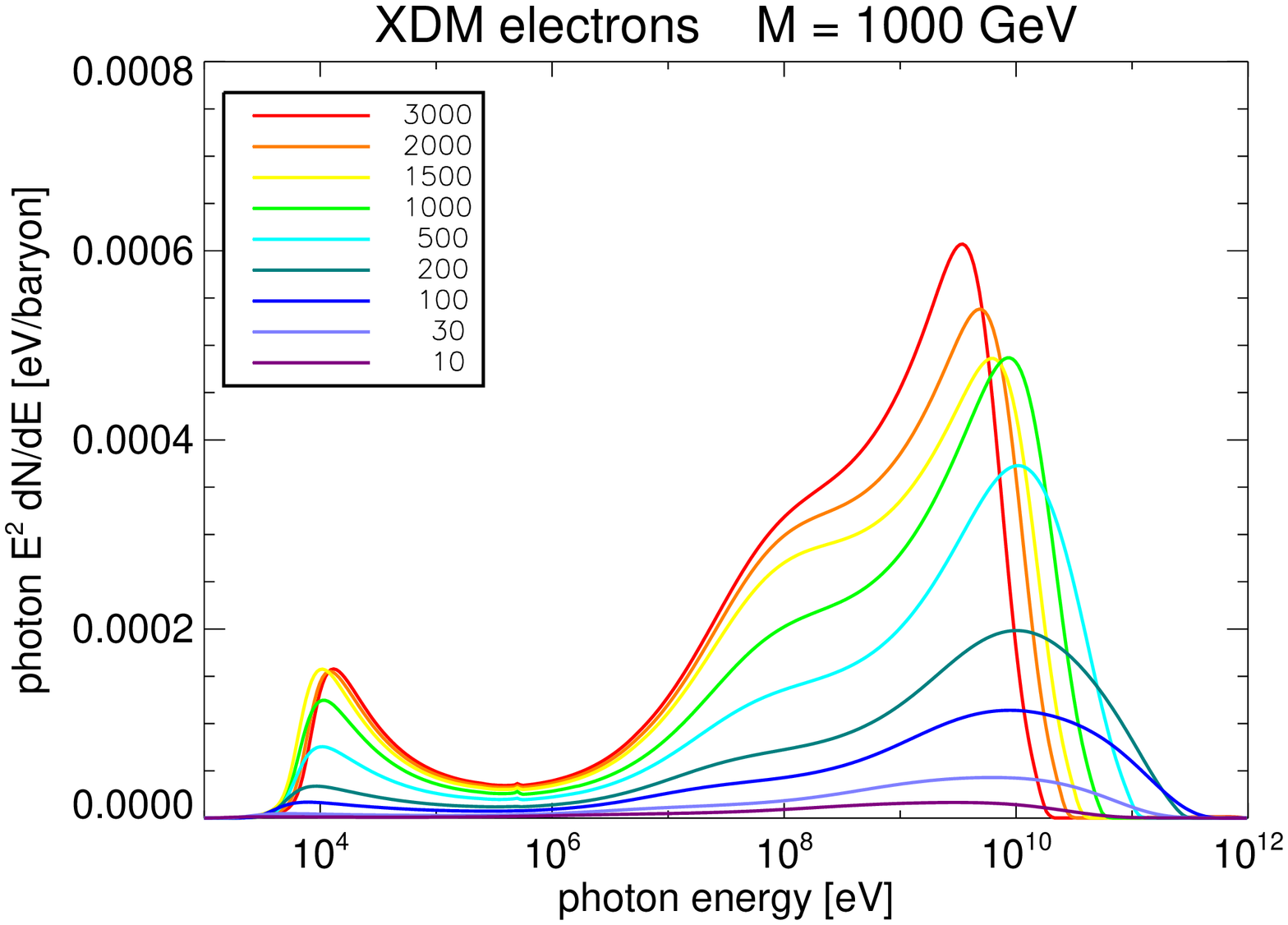}\hskip 0.2in
\includegraphics[width=3.5in]{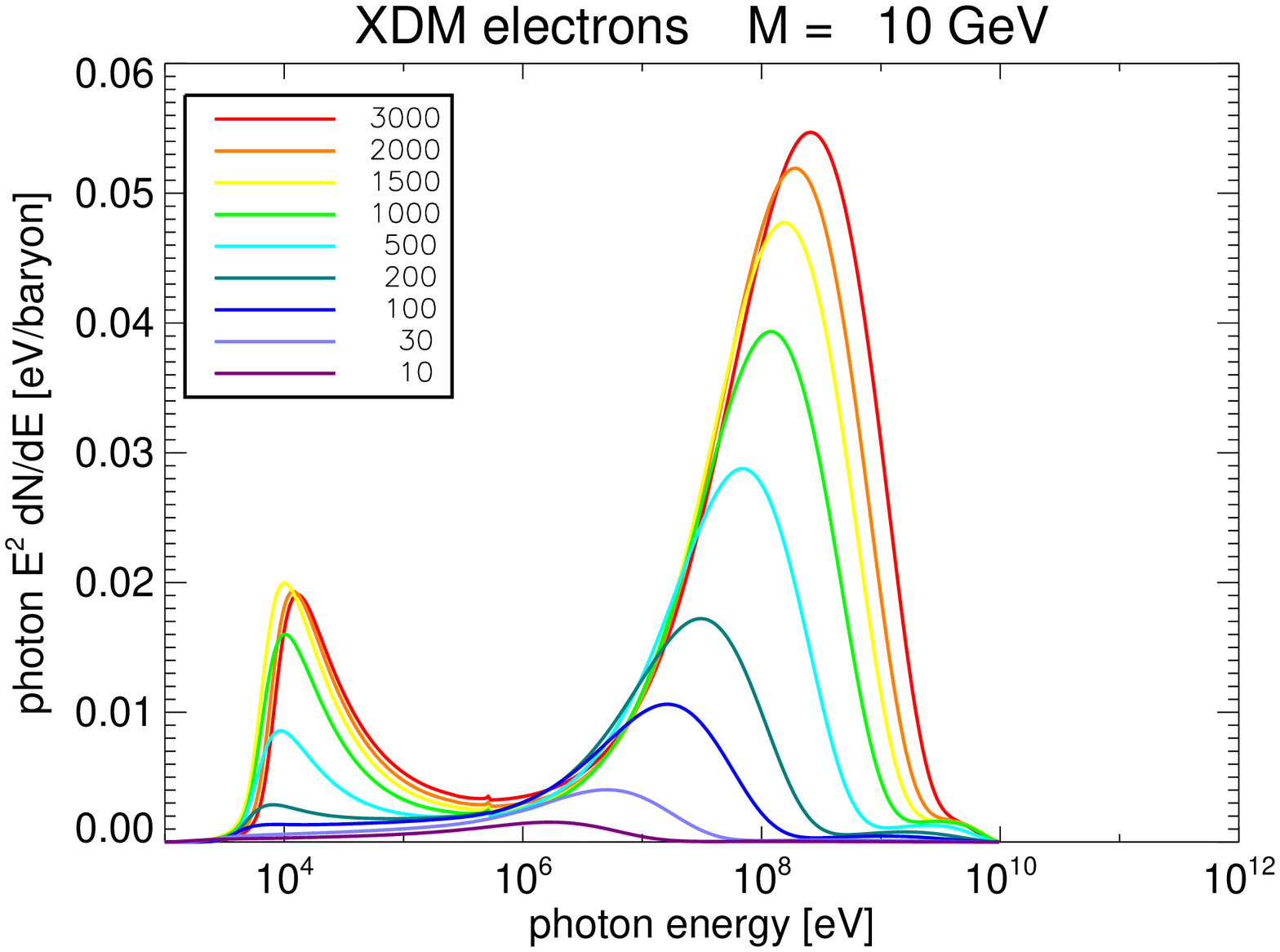}
\caption{\label{fig:photspec} 
The photon spectrum as a function of energy at several redshifts for 
$M_\mathrm{DM} = 1000$ GeV (\emph{top}) and $M_\mathrm{DM} = 10$ GeV (\emph{bottom}), 
for $\chi \chi \rightarrow \phi \phi$ followed by $\phi \rightarrow e^+ e^-$, 
with $m_\phi = 1$ GeV.}
\end{figure}

The evolution of the photon spectrum is shown for two sample models in Fig. \ref{fig:photspec}. The effect of the semi-transparent windows discussed previously is clear, with large peaks in the spectra at $\sim 10^4$ and $\sim 10^8-10^{10}$ eV. The gap between the peaks is due to Compton downscattering rapidly depleting the photon spectrum in this energy range, while above and below the semi-transparent windows, pair production on the CMB and photoionization, respectively, dominate. The edge from positron annihilation is visible at $511$ keV.

\section{Energy deposited to the IGM} 
\label{sec:fcalc}

\subsection{The efficiency factor $f(z)$}

The power per hydrogen atom injected by DM annihilation is frequently written in the form,
\begin{equation}
\epsilon_{DM} = 2 f M_\mathrm{DM} \left( \frac{\sv n_{\mathrm{DM},0}^2}{n_{\mathrm{H},0}} \right)
  (1+z)^3 
\label{eq:epsilon1}
\end{equation}
where $n_{\mathrm{DM},0}$ and $n_{\mathrm{H},0}$ are the WIMP and H
number densities at $z=0$, respectively. In the on-the-spot approximation, $f$ is just the efficiency with which the WIMP rest mass energy liberated by annihilation is injected into the IGM \footnote{In \cite{Padmanabhan:2005es} we
  absorbed the factor of 2 into the definition of $f$.}. Note that $f$ is defined in terms of the WIMP mass, not the total energy of the electrons produced by annihilation: a large branching ratio to neutrinos, for example, results in a smaller value for $f$.
  
In our previous paper on this topic \cite{Padmanabhan:2005es}, we made the simplifying assumption that $f$ and
$\sv$ were independent of redshift, an approach that has also been adopted by other authors \cite{Furlanetto:2006wp, Galli:2009zc}. The Sommerfeld enhancement can cause $\sv$ to vary with redshift, but as discussed previously, in the redshift range relevant to this problem we can reasonably assume that the enhancement is saturated (although our numerical code could trivially accommodate a time-dependent $\sv$),
\begin{equation}
\sv = S_{\mathrm{max}}\sv_\mathrm{fo} \,.
\end{equation}
Here $\sv_\mathrm{fo}$ is the usual thermal relic freeze-out cross section of $\sim 3 \times 10^{-26}$ cm$^3$/s, and $S_\mathrm{max}$ is the saturated Sommerfeld enhancement.

Our detailed numerical computation of the energy deposition allows us to go beyond assuming a constant $f$, for the models of interest. Because of the
changing transparency window (Fig. \ref{fig:transparencywindow}) the rate at which the photons' energy is absorbed by the IGM varies with $z$ and with WIMP model, even in the on-the-spot approximation. Where the on-the-spot approximation breaks down, the delayed absorption of annihilation energy injected at earlier times can also alter the effective $f(z)$ profile. To some degree, these effects may cancel each other out: the universe becomes more transparent at low redshifts and this reduces deposition efficiency, but there are more photons present than in the on-the-spot approximation, due to photons injected at higher redshift that have not yet completely cooled.

In the absence of the on-the-spot approximation, the physical meaning of $f$ as an efficiency factor is not so clear, but a slight extension of Eq. \ref{eq:epsilon1} is still a very useful parametrization. We write,
\begin{eqnarray}
\epsilon_{DM}(z) & = & 2 f(z) \frac{S_{\mathrm{max}}}{M_\mathrm{DM}} \left(\frac{\sv_\mathrm{fo}\, \rho_{\mathrm{DM},0}^2}{n_{\mathrm{H},0}} \right)
  (1+z)^3 \\
& = & 2 f(z) S_{\mathrm{max}} \left(\frac{100 \GeV}{M_\mathrm{DM}}\right) (1+z)^3 \nonumber \\
& & \times 2\times 10^{-24} \,\mathrm{eV/s/H}
\, .
\end{eqnarray}
It is this energy injection that the CMB data constrain. Note that the injected energy is inversely proportional to the particle mass; more massive
particles inject \emph{less} energy into the IGM.

\subsection{DM annihilation channels}
\label{subsec:models}

As discussed previously, recent cosmic-ray anomalies have motivated models of WIMP annihilation to leptons with a large cross section. We compute $f(z)$ for a WIMP annihilating to lepton pairs and charged pions, both directly and via a new GeV-scale state (annihilation channels of the latter type are denoted ``XDM''). As a benchmark, the mass of the new light state is taken to be 1 GeV for electron, muon and pion final states, and 4 GeV for taus: however, because of the large mass hierarchy between the WIMP and the GeV-scale state, the spectrum of the SM annihilation products is nearly independent of this choice of parameter. We investigate a range of WIMP masses for these annihilation channels, including the mass/channel combinations fitted to PAMELA and ATIC data in \cite{Cholis:2008wq}. We also compute $f(z)$ for WIMPs annihilating to pairs of SM particles, for benchmark WIMP masses of 200 GeV and 1000 GeV. 

We note that at sufficiently high $M_\mathrm{DM}$, the photons from annihilation are injected at energies where pair production on the CMB is extremely rapid, and so the spectrum of produced photons is entirely determined by the pair production + ICS cascade. Consequently, around the redshift of last scattering, $f(z)$ becomes essentially independent of the WIMP mass at $M_\mathrm{DM} \ga 1$ TeV.

\begin{figure*}
\begin{center}
\includegraphics[width=.45\textwidth]{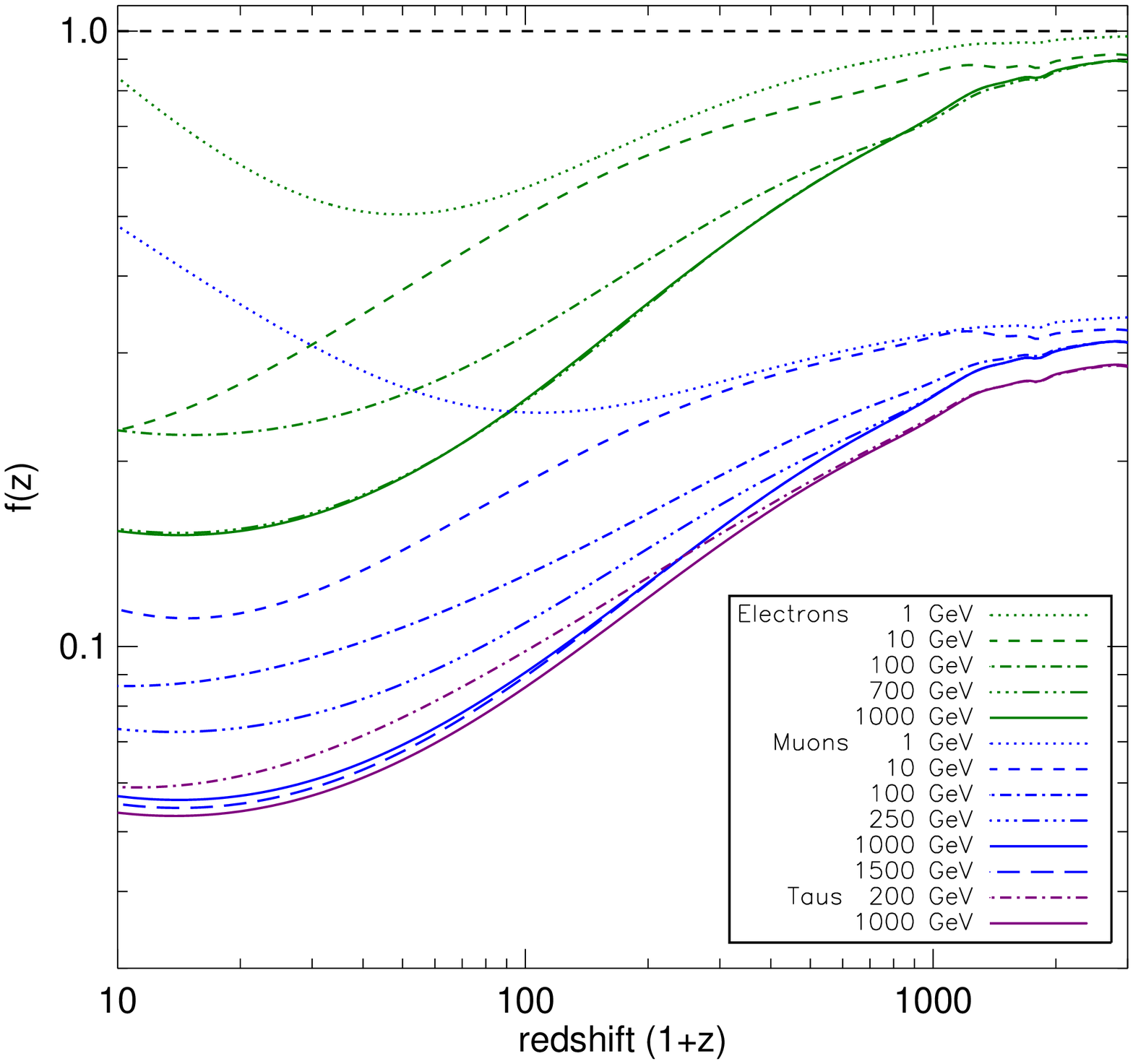}\hskip 0.2in
\includegraphics[width=.45\textwidth]{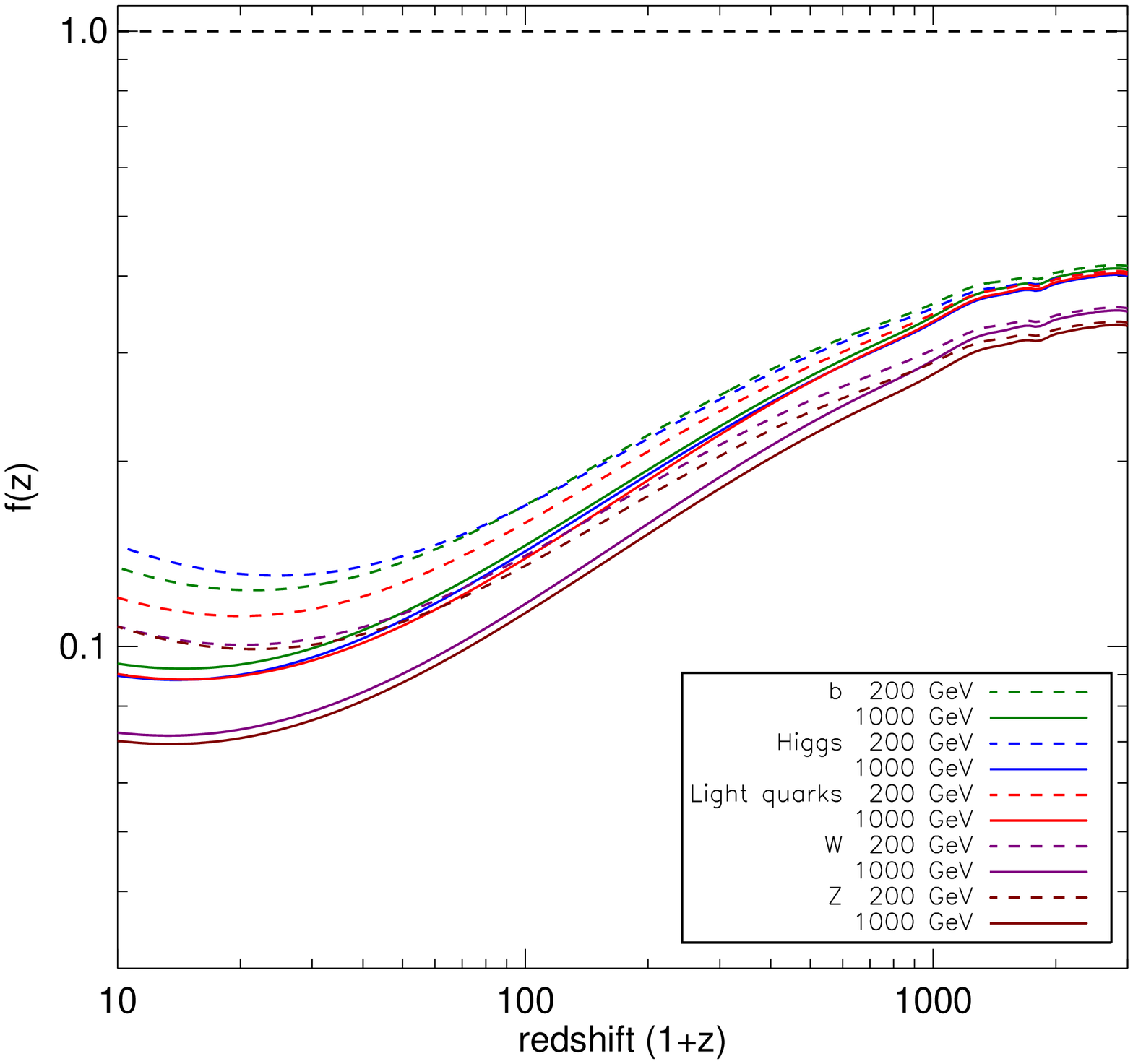}\\
\includegraphics[width=.45\textwidth]{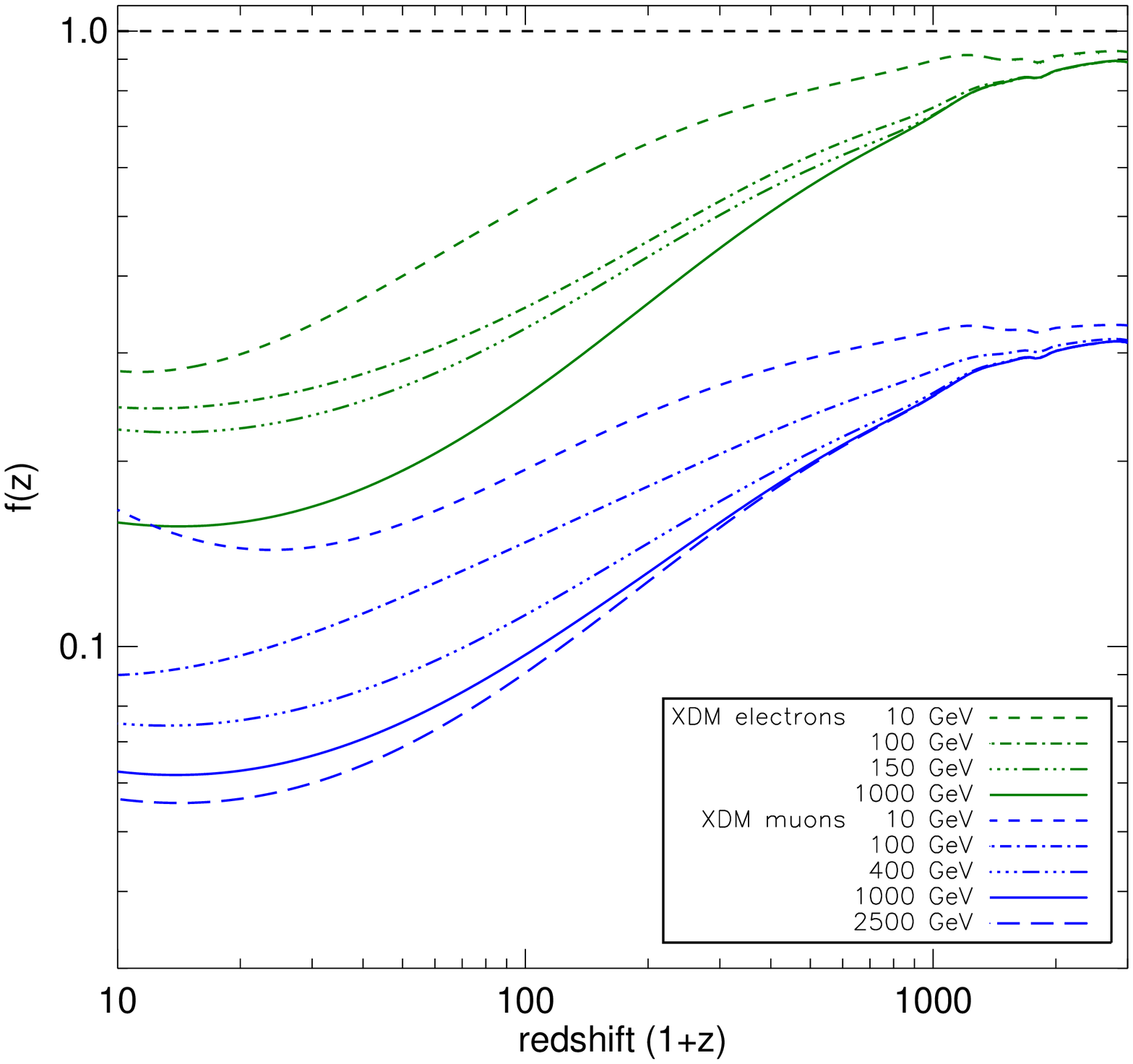}\hskip 0.2in
\includegraphics[width=.45\textwidth]{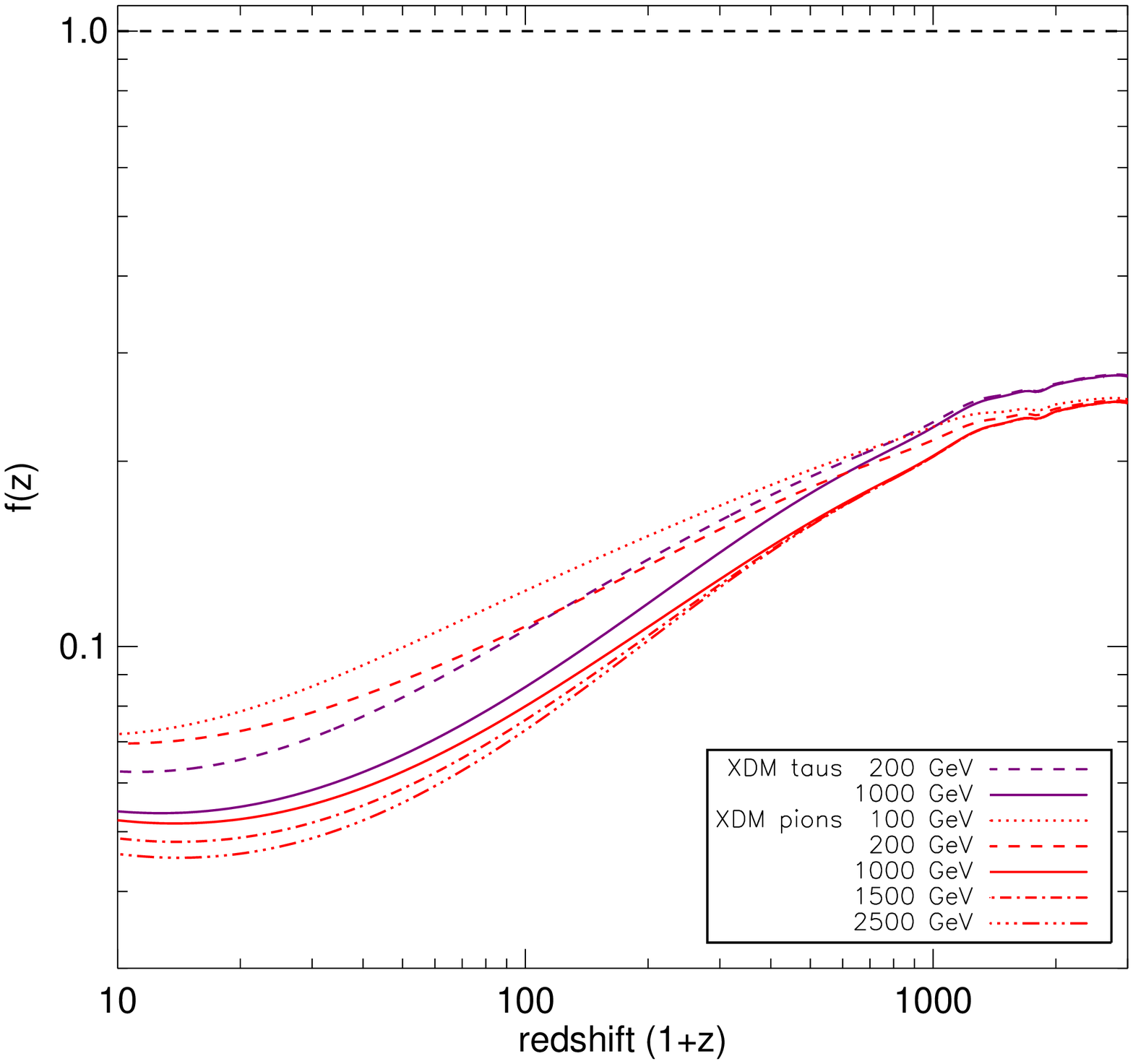}
\end{center}
\caption{\label{fig:fracdeposited}
The ``deposited power fraction'' $f(z)$ is the ratio of the power
deposited in the gas (in the form of ionizations, excitations, and
heating) to the mass energy liberated by WIMP annihilations. For electron channels, $f(z) \sim 1$ at high $z$, but other channels lose some fraction of their power to neutrinos, protons and neutrons.
\emph{Upper left panel}: direct annihilation to SM leptons. \emph{Upper right panel}: direct annihilation to non-leptonic SM states (``light quarks'' corresponds to 50 $\%$ annihilation to u quarks, 50 $\%$ to d quarks). \emph{Lower left panel}: XDM-type models with annihilation through an intermediate 1 GeV state to electrons and muons. \emph{Lower right panel}: XDM-type models with annihilation through an intermediate 1 GeV state to charged pions, and through an intermediate 4 GeV state to taus. 
The legend indicates the annihilation channel and the WIMP mass. The kink
around $z=1700$ is an artifact of an approximation made in
\texttt{RECFAST} and has no impact on our results.}
\end{figure*}

Taking linear combinations of these channels allows $f(z)$ to be computed accurately for a wide range of DM models. In particular, models where the DM does not originate in thermal equilibrium can have large annihilation cross sections, and so can be strongly constrained by the CMB once $f(z)$ is known precisely. The indirect detection prospects of such nonthermal neutralino models have been investigated in \cite{Moroi:1999zb,Kane:2001fz,Nagai:2008se,Grajek:2008jb}. Recently, a 200 GeV wino with annihilation cross section $\sigmav = 2 \times 10^{-24}$ cm$^3$/s has been proposed as an explanation for the positron excess observed by PAMELA \cite{Grajek:2008pg}. In this model the neutralino annihilates predominantly to $W$ bosons, and $f(z)$ and the impact of DM annihilation on the CMB can therefore be immediately computed from our results.
We calculate the spectra of $e^+ e^-$ and photons resulting from these annihilation channels (including final state radiation) using \texttt{PYTHIA}. Fig. \ref{fig:fracdeposited} displays $f(z)$ as a function of redshift for the various WIMP masses and annihilation channels.

In Appendix \ref{app:fitparams} we present accurate fitting functions for $f(z)$ over the redshift range $z=300-1200$, for all the annihilation channels under consideration. Our choice of redshift range is dictated by the impact on the visibility function: above $z \sim 1200$ or below $z \sim 500$, changes to $f(z)$ have minimal impact on the last scattering surface (see for example Fig. 3 in \cite{Padmanabhan:2005es}). Less precise fits are given for higher and lower redshifts. Note that these fitting functions do not apply to models of decaying dark matter: while our numerical approach can be applied to decaying WIMP models with only trivial changes, we defer such an analysis to future work.

\section{CMB Constraints}
\label{sec:cmbconstraints}

CMB constraints on the energy injection from DM annihilation, assuming a constant $f(z)$, have been previously calculated in \cite{Padmanabhan:2005es} and updated in \cite{Galli:2009zc}. Using the approximate values for $f(z)$ given in Table \ref{tab:fitparams}, we can now relate these constraints to specific DM models. A full analysis of the CMB constraints would take into account the variation of $f(z)$ with redshift, but since $f(z)$ is slowly varying in the redshift range of interest, we employ the approximation of a constant $f$ to estimate the degree to which the CMB can constrain models of interest, and defer a more complete analysis to future work.

\begin{figure*}[ht]
\begin{center}
\includegraphics[width=.32\textwidth]{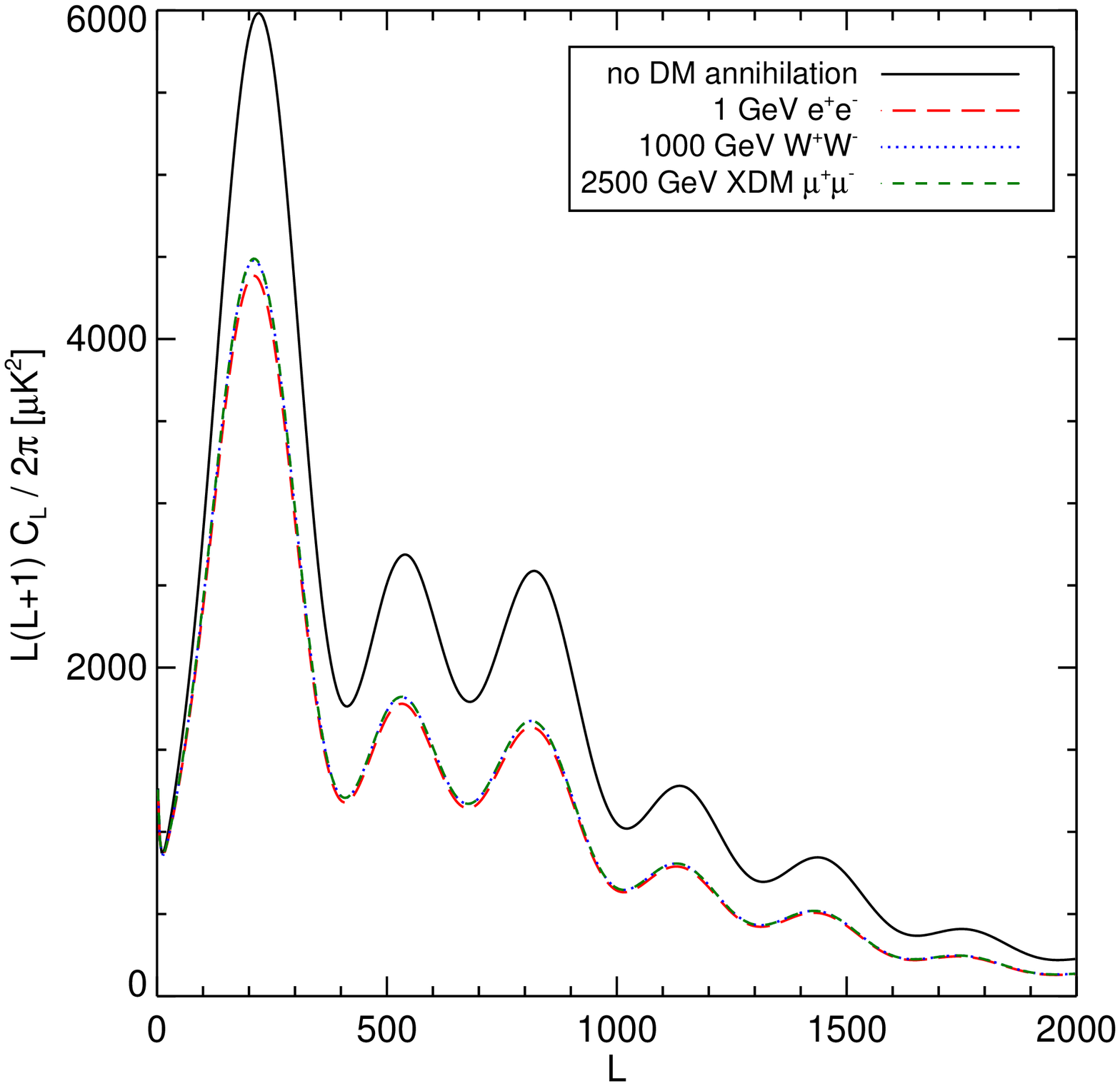}
\includegraphics[width=.32\textwidth]{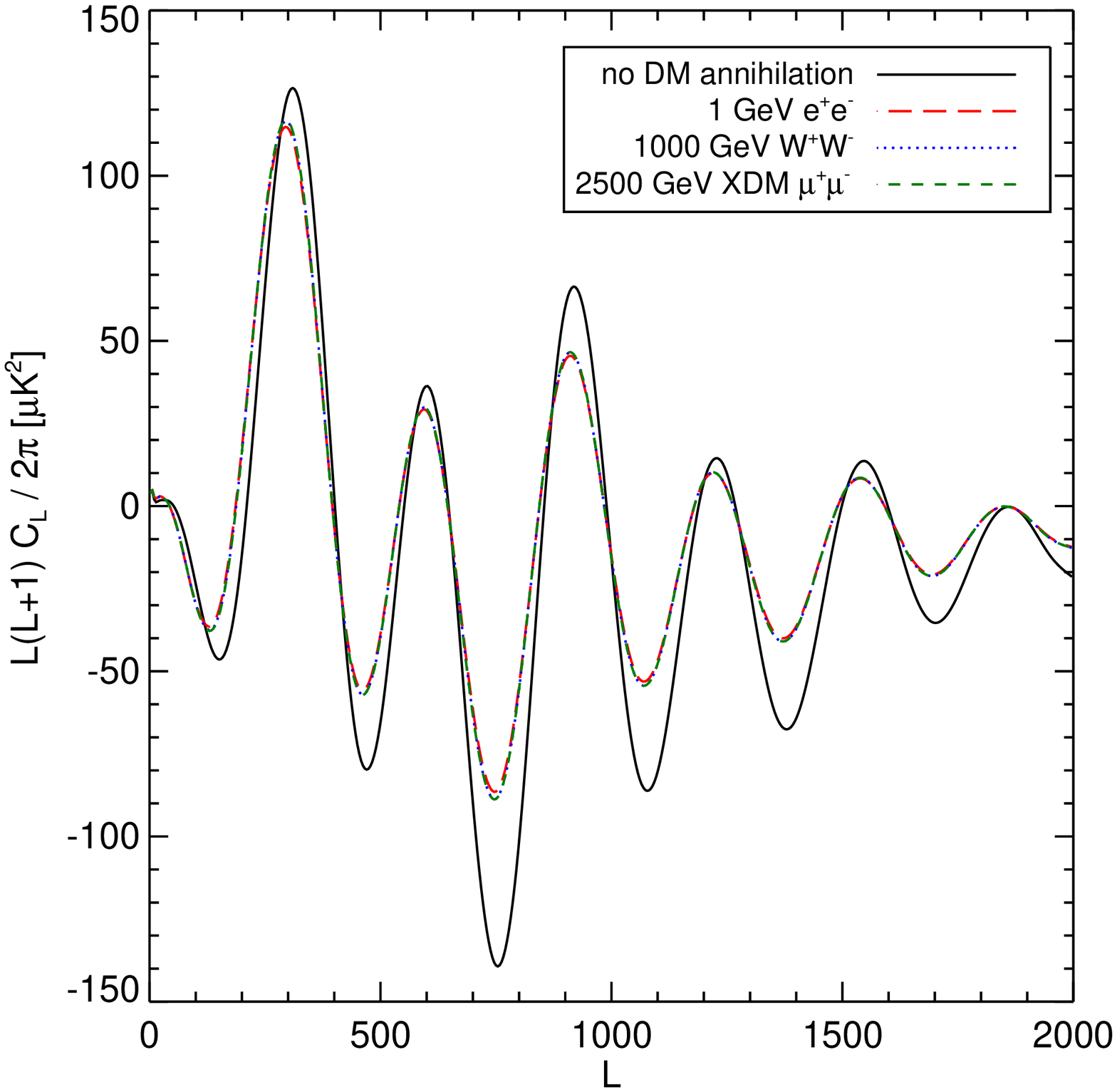}
\includegraphics[width=.32\textwidth]{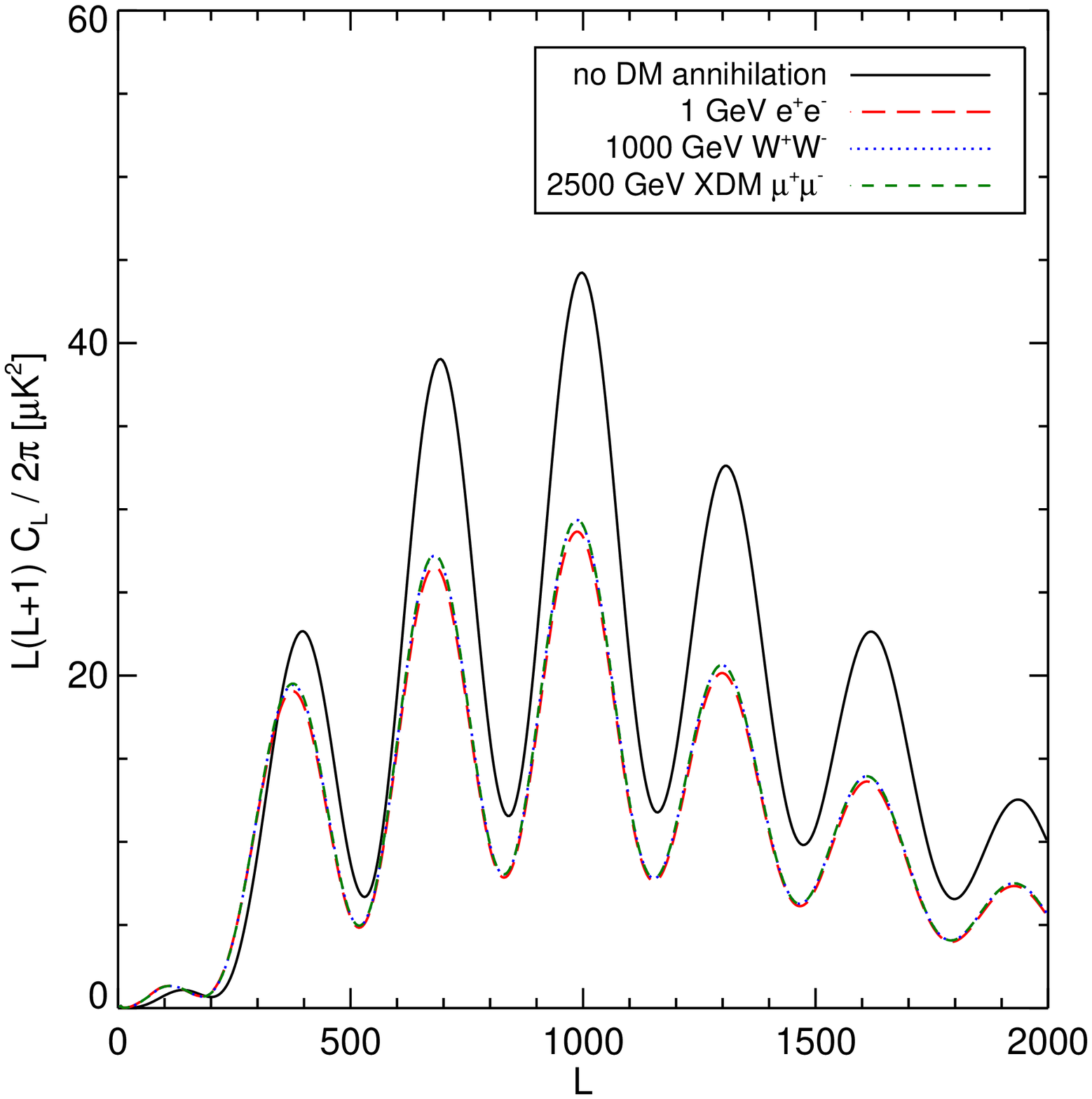}\\
\end{center}
\caption{\label{fig:cmbplots}
  CMB power spectra for three different DM annihilation models, with power injection normalized to that of a 1 GeV WIMP with thermal relic cross section and $f=1$, compared
  to a baseline model with no DM annihilation.  The models give similar
  results for the TT (\emph{left}), TE (\emph{middle}), and EE
  (\emph{right}) power spectra.  This suggests that the
  CMB is sensitive to only one parameter, the average power injected
  around recombination. All curves employ the WMAP5 fiducial cosmology: the effects of DM annihilation can be compensated to a large degree by adjusting $n_s$ and $\sigma_8$ \cite{Padmanabhan:2005es}. }
\end{figure*}

To a large extent, the effect of DM annihilation on the CMB depends only on the average of $f(z)$ in the redshift range of interest $z \sim 800- 1000$, not the details of $f(z)$. We employed the public codes \texttt{RECFAST} and \texttt{CAMB} \cite{2000ApJ...538..473L} to derive the effect of DM annihilation on the TT, TE and EE angular power spectra, for several different $f(z)$ profiles (with the widest possible differences in mass and annihilation channel), normalizing the annihilation cross sections so that the deposited power $\epsilon(z)$ would be the same in all cases if $f(z)$ was replaced by its mean value between $z=800-1000$. Fig. \ref{fig:cmbplots} demonstrates the results: even in an extreme case where the cross section is so large as to be strongly excluded by current constraints, the differences in the spectra due to differing $f(z)$ profiles are small.

\subsection{Models fitting cosmic-ray excesses}
\begin{figure}[h]
\includegraphics[width=3.5in]{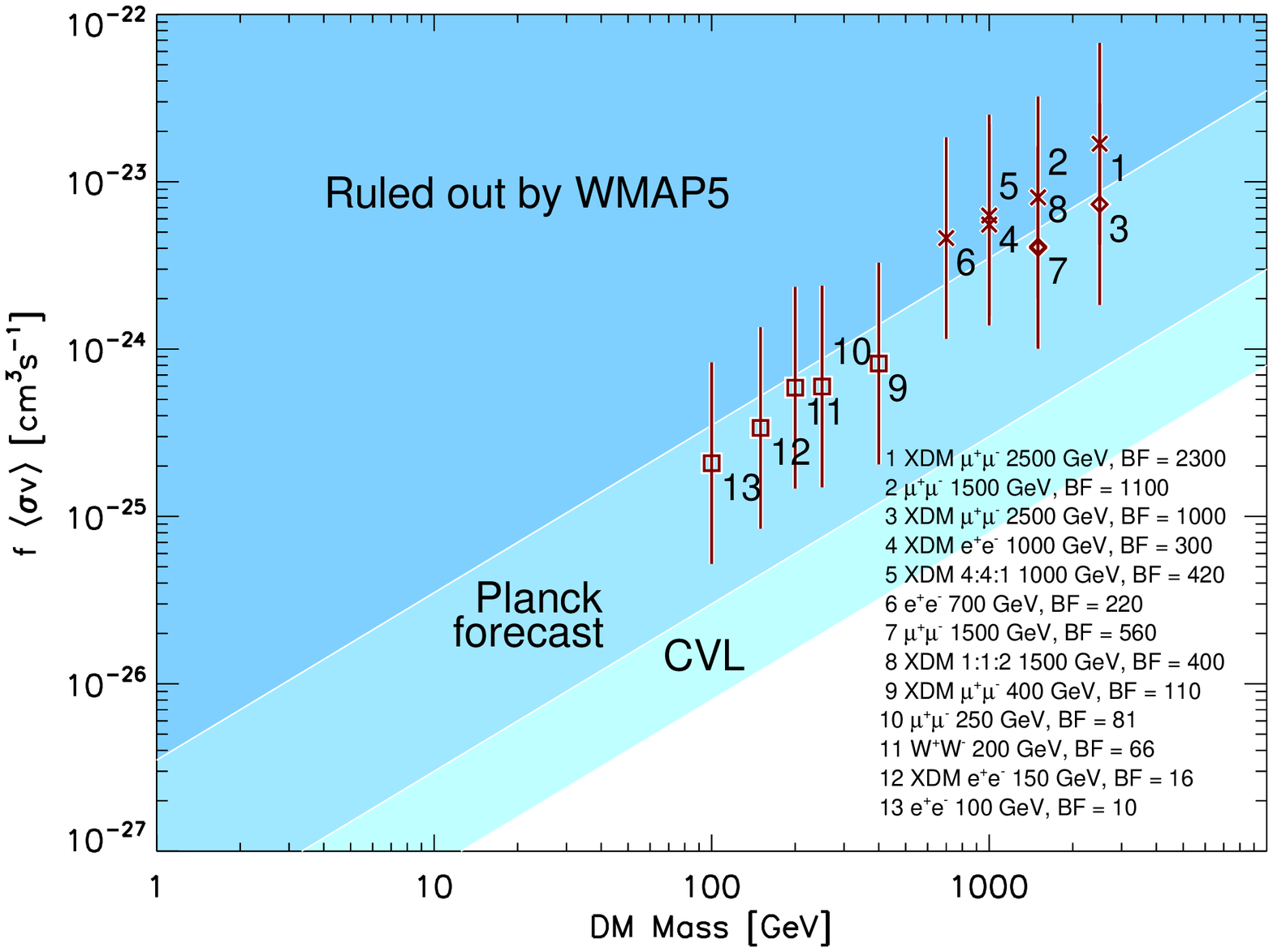}
\caption{\label{fig:constraints}
Constraints on the annihilation cross-section $\sigmav$ the efficiency factor
$f$. The dark blue area is excluded by WMAP5
data at $95 \%$ confidence, whereas the lighter blue area shows the region of parameter space that will be probed by Planck.
The cyan area is the zone that can ultimately be explored by
a cosmic variance limited experiment with angular resolution
comparable to Planck. Constraints are taken from \cite{Galli:2009zc} (Fig. 4).
The data points indicate the positions of models which fit the observed cosmic-ray excesses, as fitted in \cite{Cholis:2008wq, Grajek:2008pg}. \emph{Squares:} PAMELA only. \emph{Diamonds:} PAMELA and Fermi. \emph{Crosses:} PAMELA and ATIC. Error bars indicate the factor-of-4 uncertainty in the required boost factor due to uncertainties in the local dark matter density (any substructure contributions are not taken into account).  For models labeled by ``XDM'' followed by a ratio, the annihilation is through an XDM intermediate light state to electrons, muons and pions in the given ratio (e.g. ``XDM 4:4:1'' corresponds to 4:4:1 annihilation to $e^+ e^-$, $\mu^+ \mu^-$ and $\pi^+ \pi^-$).}
\end{figure}

We focus here on models which fit the cosmic-ray excesses measured by PAMELA, and in the case of higher-mass WIMPs, also ATIC or Fermi. Boost factors and WIMP masses are taken from \cite{Cholis:2008wq} for the leptonic and XDM channels, and from \cite{Grajek:2008pg} for annihilation to $W$ bosons. Fig. \ref{fig:constraints} displays the WMAP5 constraints on these models, and the region of parameter space that will be probed by Planck.

In all cases, the models which fit the cosmic-ray excesses are close to being ruled out by WMAP5, at 95 $\%$ confidence. The tension is greater for models which fit the ATIC excess, where the boost factors given in \cite{Cholis:2008wq} are already excluded. However, this result does not  rule out these DM models as explanations for the ATIC excess, due to astrophysical uncertainties in the required boost factor. For example, the local DM density is only known to within a factor of $\sim 2$ (which is then squared to determine the annihilation rate), and density enhancements from local substructure could also contribute an $\mathcal{O}$(1) boost to the cosmic-ray flux. The excess measured by Fermi requires generically smaller boost factors than ATIC, by a factor of $\sim 2-3$: such models are not ruled out by WMAP5 even without taking into account astrophysical uncertainties, but will be constrained by Planck.

The degree of uniformity between the models should not be surprising, despite the wide range of masses and boost factors. The variations in $f(z)$ between different channels arise in large part from the energy carried away by annihilation products other than photons and electrons -- but these annihilation products also do not contribute to the cosmic-ray excesses measured at ATIC and PAMELA. The cosmic-ray excesses are more sensitive measures of the high-energy spectrum of the annihilation products than the CMB, whereas the CMB is sensitive to soft photons and electrons which may be absorbed into the background in cosmic-ray measurements, but to a first approximation both measurements are simply probing the total power in electrons (at least when the power in photons produced by annihilation is small).

\subsection{Implications for Sommerfeld-enhanced DM annihilation}

As described in the Introduction, the CMB has the potential to act as an especially sensitive probe of DM models with Sommerfeld-enhanced annihilation. The simplest example of the Sommerfeld enhancement with a massive mediator is the case of WIMPs interacting via a Yukawa potential. More complicated models can contain small mass splittings among the dark sector particles, and multiple light force carriers (e.g. \cite{ArkaniHamed:2008qn}), but in this work we will consider only the simplest case.

If the dark matter particle couples to a scalar mediator $\phi$ with coupling strength $\lambda$, then the enhancement is solely determined by the dimensionless parameters,
\begin{equation} \epsilon_v= \frac{(v/c)}{\alpha}, \,\, \epsilon_\phi = \frac{m_{\phi}}{\alpha M_{\mathrm{DM}}},
\end{equation}
where $\alpha = \lambda^2 / 4 \pi$. In the limit where the $\phi$ mass goes to zero ($\epsilon_\phi \rightarrow 0$), the  enhancement to the annihilation cross section -- denoted $S$ -- can be determined analytically, and $S \sim \pi/\epsilon_v$ at low velocities. For nonzero $\epsilon_\phi$,  there are
two important qualitative differences. The first is that the
Sommerfeld enhancement saturates at low velocity--the attractive
force has a finite range, and this limits how large the enhancement
can become. Once the deBroglie wavelength of the
particle $(M_\mathrm{DM}v)^{-1}$ exceeds the range of the interaction
$m^{-1}_\phi$, or equivalently once $\epsilon_v$ drops beneath
$\epsilon_\phi$, the Sommerfeld enhancement saturates at $S \sim
\frac{1}{\epsilon_\phi}$ \cite{ArkaniHamed:2008qn}. The second effect is that for specific
values of  $\epsilon_\phi$, resonances occur where the enhancement scales as $\sim 1/\epsilon_v^2$ instead of $\sim 1/\epsilon_v$, potentially increasing the enhancement factor by several orders of magnitude. In the resonant case the velocity at which the enhancement saturates is also smaller than in the non-resonant case (for the same value of $\epsilon_\phi$).

\subsubsection{Saturation of the enhancement}
At first glance it might appear that our calculation would not apply to Sommerfeld-enhanced models, due to the variation of the enhancement with velocity, since we have assumed a constant $\sigmav$ with respect to $z$. However, for models which are not already ruled out by WMAP5 constraints, either the enhancement must be saturated over the redshift range in question ($z \sim 100-4000$), or $\alpha$ or $f(z)$ must be extremely small -- in which case the model could not explain the cosmic-ray anomalies described in the Introduction. For the models of greatest interest, the enhancement $S$ thus provides a constant boost factor to the annihilation cross section at $z \sim 1000$, and our constraints apply directly.

At redshift $z$, the CMB temperature is $\sim 2.35 \times 10^{-4} (1+z)$ eV. This places an upper bound on the temperature of the DM: however, after kinetic decoupling the DM temperature evolves adiabatically as $T \propto z^2$, and thus the WIMPs can be much colder than the photon temperature. \cite{Galli:2009zc} suggests $v/c \sim 10^{-8}$ at $z \sim 1000$ for a 100 GeV WIMP.

If the enhancement is still unsaturated at such low velocities, then the force carrier must be extremely light compared to the WIMP mass. For the models recently proposed in the literature \cite{Cirelli:2007xd, ArkaniHamed:2008qn, Nomura:2008ru, Baumgart:2009tn}, the enhancement has always saturated by this point as the force carriers are much heavier than $10^{-8} M_\mathrm{DM}$. Other constraints on models with very low-mass mediators also exist: as one example, a $1/v$ enhancement which saturates at too low a velocity can also cause runaway annihilations in the first DM halos at the onset of structure formation \cite{Kamionkowski:2008gj}. Furthermore, as shown in Fig. \ref{fig:constraints}, models which fit the recently observed cosmic-ray anomalies are already close to being ruled out by WMAP5. If the Sommerfeld enhancement in such models has not saturated by $(v/c) \sim 10^{-8}$, this implies an effective cross section at recombination $\sim 4-5$ orders of magnitude higher than in the present-day Galactic halo. Such models are therefore \emph{strongly} excluded by WMAP5. Similarly, if the WIMP annihilates to the same particle which mediates the Sommerfeld enhancement, then in order for the enhancement to evade the constraints in Fig. \ref{fig:constraints}, the coupling $\alpha$ between the WIMP and the force carrier must be extremely small -- reducing the annihilation cross section at freeze-out to unacceptable levels for a thermal relic. Thus for a broad range of well motivated models, it is self-consistent to assume that the Sommerfeld enhancement is saturated for the redshift range of interest ($z \sim 100-4000$).  

We can write the 95 $\%$ confidence limits from WMAP5 in terms of constraints on the total cross section,
\begin{equation} \sigmav_\mathrm{saturated} < \frac{3.6 \times 10^{-24} \mathrm{cm}^3/\mathrm{s}}{f}\left( \frac{M_\mathrm{DM} c^2}{\mathrm{1 TeV}} \right), \end{equation} 
or as constraints on the maximum saturated enhancement, relative to the thermal relic cross section $\sigmav = 3 \times 10^{-26}$ cm$^3$/s,
\begin{equation} S_\mathrm{max} < \frac{120}{f}\left( \frac{M_\mathrm{DM} c^2}{\mathrm{1 TeV}}\right). \label{eq:enhanceconstraints} \end{equation} 
In both cases values of $f$ for the different channels are given in Table \ref{tab:fitparams}.

These results directly limit the maximum boost factor possible from substructure, in Sommerfeld-enhanced models. There has recently been considerable interest in possible annihilation signals from dark matter subhalos, where the DM velocity dispersion is reduced and the Sommerfeld-enhanced cross section is boosted (e.g. \cite{Lattanzi:2008qa,Yuan:2009bb, Bovy:2009zs, Kuhlen:2009is}). However, the saturated cross section cannot be much larger than that required to fit the cosmic ray anomalies, so for models which fit the cosmic ray anomalies, the lower velocity dispersion in subhalos will not result in a higher annihilation cross section. 

\subsubsection{Sommerfeld-enhanced models fitting cosmic ray excesses}

In Sommerfeld-enhanced models which produce the observed excesses in $e^+ e^-$ cosmic rays, the saturation of the enhancement is even more constrained than in the general case. Since the cross sections required to fit the cosmic ray anomalies are already nearly excluded by WMAP5, as shown in Fig. \ref{fig:constraints}, the enhancement must already be close to saturation at $v \sim 150$ km/s ($5 \times 10^{-4} c$), the estimated local WIMP velocity dispersion. Astrophysical uncertainties -- in the propagation of cosmic rays, the local dark matter density and the local velocity distribution -- can weaken this constraint by a factor of a few. 

In the case of the simplest Sommerfeld enhancement scenario, where the light force carrier generates a Yukawa potential between the WIMPs, this saturation requirement disfavors models with $m_\phi \ll 10^{-4} M_\mathrm{DM}$. Stronger constraints on the force carrier mass may be possible in the case of resonant enhancement, where $\epsilon_v < \epsilon_\phi$ at saturation.

\subsubsection{Annihilation through the force carrier}
A related but independent constraint on the force carrier mass occurs for models where the WIMP annihilates to the same particle that mediates the Sommerfeld enhancement, provided the WIMP is a thermal relic and constitutes 100 $\%$ of the dark matter \footnote{We thank Adam Ritz and Brian Batell for pointing out this argument.}. In this case the WIMP mass and its coupling to the light force carrier satisfy the relation \cite{Pospelov:2008jd},
\begin{equation} \alpha \approx 10^{-2} \times \frac{M_\mathrm{DM} c^2}{270~\mathrm{GeV}}, \end{equation}
since the freeze-out cross section for WIMP annihilation to the light force carrier is $\sigmav \sim \pi\alpha^2 / M_\mathrm{DM}^2$. But in the case of non-resonant Sommerfeld enhancement with behavior similar to the simple Yukawa case, the saturated enhancement is given by,
\begin{equation} S_\mathrm{max} \sim \frac{\alpha}{m_\phi / M_\mathrm{DM}} \sim 10^{-2} \times \frac{(M_\mathrm{DM} c^2)^2}{(270~\mathrm{GeV}) m_\phi c^2}. \end{equation}
Applying our constraints on $S_\mathrm{max}$ in the form given in Eq. \ref{eq:enhanceconstraints}, it follows that,
\begin{equation} \frac{m_\phi}{M_\mathrm{DM}} \ga 10^{-4} \times \left( \frac{f}{0.3} \right). \label{eq:massconstraint} \end{equation}
In this case the constraint does not depend on the cosmic-ray data and thus there are no uncertainties from present-day Galactic astrophysics. However, $\mathcal{O}$(1) model-dependent factors in the saturated enhancement and the annihilation cross section at freeze-out may modify Eq. \ref{eq:massconstraint}, and in the case of resonant enhancement, the constraint can become significantly stronger (by a factor $S_\mathrm{max} \epsilon_\phi$): these factors can be computed given a specific DM model, and such a calculation is necessary to make the limit precise. As previously, though, force carriers with masses $\ll 10^{-4} M_\mathrm{DM}$ are disfavored. For WIMPs with masses of order a few TeV, as suggested by fits to the Fermi and H.E.S.S. data, this constraint may be sufficient to rule out force carriers with masses below the muon threshold, in a wide range of models.

\section{Conclusions}

The CMB provides a robust constraint on models of dark matter annihilation, which is particularly relevant for models that annihilate rapidly enough today to produce the PAMELA, Fermi and/or ATIC signals.  These signals all
require an annihilation cross section larger than the thermal relic cross section,
implying either non-thermal production or some other mechanism, such
as Sommerfeld enhancement.  In either case, the implied cross section
at $z\approx 1000$ is as large as today, or possibly larger.

Extending previous work \cite{Chen:2003gz, Padmanabhan:2005es, Ripamonti:2006gq, Galli:2009zc}, we have computed the efficiency of
annihilation power deposition in the IGM for a number of models, some
tuned to fit the recently measured cosmic-ray anomalies, and some with more general
annihilation channels.  We have improved previous calculations by finding and correcting some errors in the photon cooling processes, and by treating the redshifting and downscattering of photons in the semi-transparent windows in detail.  Numerical convergence on this calculation
indicates it is correct at the $\sim 1\%$ level, 
an accuracy far exceeding that of any possible constraints on local DM annihilation, as they depend on e.g. the local DM
density squared, which is uncertain.

Given the state of the current data, it is adequate to approximate the
efficiency of energy deposition, $f(z)$, with an average over the redshift range of interest.  Future improvements in the data, however, may
demand a more detailed calculation.  Therefore we provide fitting
functions for 40 models of interest in Appendix \ref{app:fitparams}.

WMAP5 already constrains Sommerfeld-enhanced models that have been proposed to explain the recently observed cosmic-ray excesses, requiring that the saturated annihilation cross section does not greatly exceed the cross section in the neighborhood of the Earth. This result argues against large DM annihilation signals from substructure, in the present-day Galaxy. In the simplest case where the force carrier is a scalar generating a Yukawa potential, and either the model produces the observed cosmic-ray signals or the DM is a thermal relic annihilating to the same particle that mediates the Sommerfeld enhancement, masses for the force carrier much less than $10^{-4} M_\mathrm{DM}$ are disfavored. 

For a local DM density of $0.3$ GeV/cm$^3$, standard assumptions for cosmic-ray propagation \cite{Cholis:2008wq}, and no contribution to the annihilation flux from substructure, there is mild tension (at the factor of 2 level) between the WMAP5 95 $\%$ confidence limits and models fitting the excess observed by ATIC. However, due to $\mathcal{O}$(1) uncertainties in the local density and substructure contribution, these explanations for the ATIC excess cannot be excluded by WMAP5. Models fitting the excesses measured by PAMELA and Fermi are consistent with the WMAP5 constraints, for standard astrophysical assumptions. However, all of these models will be rigorously tested by the recently launched Planck experiment. 

We wish to acknowledge helpful conversations with Nima Arkani-Hamed, Brian Batell, Gianfranco Bertone, Xuelei Chen, Ilias Cholis, Lisa Goodenough, Marc Kamionkowski, Priya Natarajan, Adam Ritz, Philip Schuster,  Natalia Toro, and Neal Weiner.  We are especially grateful to Neal Weiner for providing the
\texttt{PYTHIA} outputs used in this work.  
DPF is partially supported by NASA LTSA grant NAG5-12972.  NP is
supported by NASA Hubble Fellowship HST-HF-01200.01 awarded by the
Space Telescope Science Institute, which is operated by the
Association of Universities for Research in Astronomy, Inc., for NASA,
under contract NAS 5-26555.  NP is also supported by an LBNL
Chamberlain Fellowship.

\newpage
\onecolumngrid
\appendix

\section{Fitting functions for $f(z)$}
\label{app:fitparams}

Around the redshift of last scattering, $f(z)$ is slowly varying with respect to redshift; the approximation of a constant $f(z)$ is not too far from the truth. As described in Section \ref{sec:cmbconstraints}, we average $f(z)$ over the redshift range $z=800-1000$ to obtain an effective constant efficiency $f$, for a range of masses and annihilation channels. For applications where a more accurate approximation to $f(z)$ is required, we fit $f(z)$ to a seven-parameter function over the redshift range $z=300-1200$,
\begin{equation} f(z) = F (1+z)^\alpha  \left( \left(\frac{1+z}{z_0} \right)^\gamma + \left( \frac{1+z}{z_0} \right)^{-\gamma} \right)^\beta   \exp \left( \frac{\delta}{1 + ((1+z)/1100)^\eta} \right).  \label{eq:ffit} \end{equation}
 These fits are accurate to within $1 \% $ between $z=300-1200$ for all channels. These fits remain accurate to $< 5 \%$ between $z=170$ and $z=1470$, but outside this range they may perform very poorly.

%---------------------------------------------- TABLE -------------------------------------
\begin{table}[h]
\centering \tiny
\begin{tabular}{r|r@{\hspace{3 pt}}|@{\hspace{8 pt}}r|@{\hspace{8 pt}}c|@{\hspace{3 pt}}rrr|@{\hspace{3 pt}}rrrrrrr}

\hline
                &   DM mass &                          &  &&&  &&&&&&&      \\
        Channel &     (GeV) &     $f_\mathrm{mean}$ &     $f(z=2500)$ & $a$  &     $b$  &     $c$ &     F &     $\alpha$ & $\beta$ & $\gamma$ & $\delta$ & $\eta$ & $z_0$ \\
\hline
      Electrons & $    1 $ & $  0.92 $ & $  0.98$ & $   0.5069 $ & $  51.8802 $ & $   2.2828$ & $   0.1140 $ & $   0.4099 $ & $  -0.5634 $ & $   0.6445 $ & $   0.0043 $ & $  -5.1992 $ & $ 150.3970$ \\     
      $\chi \chi \rightarrow e^+ e^-$ & $   10 $ & $  0.84 $ & $  0.91$ & $  0.0715 $ & $   0.0078 $ & $   6.7966$ & $   0.0864 $ & $   0.4028 $ & $  -0.2453 $ & $   1.1481 $ & $   0.0488 $ & $  -4.1911 $ & $ 166.4426$\\
       & $  100 $ & $  0.69 $ & $  0.89$ & $   0.2207 $ & $  14.5754 $ & $   3.1748$ & $   0.0676 $ & $   0.3745 $ & $  -0.1973 $ & $   0.9745 $ & $   0.0682 $ & $ -13.0681 $ & $ 322.3401$ \\
       & $  700 $ & $  0.70 $ & $  0.89$ & $   0.1527 $ & $  13.3065 $ & $   2.8822$ & $   0.0841 $ & $   0.3698 $ & $  -0.5719 $ & $   0.5410 $ & $   0.0528 $ & $ -12.3998 $ & $ 663.9780$ \\
       & $ 1000 $ & $  0.70 $ & $  0.89$ & $   0.1515 $ & $  13.3421 $ & $   2.8416$ & $   0.0701 $ & $   0.3696 $ & $  -0.3077 $ & $   0.7263 $ & $   0.0469 $ & $ -12.9124 $ & $ 678.7171$ \\
      \hline
          Muons & $    1 $ & $  0.32 $ & $  0.34$ & $   0.2396 $ & $ 133.1554 $ & $   3.0272$ & $   0.0602 $ & $   0.3284 $ & $  -0.4350 $ & $   0.5484 $ & $  -0.0094 $ & $  -4.7619 $ & $  97.2662$ \\
       $\chi \chi \rightarrow \mu^+ \mu^-$  & $   10 $ & $  0.31 $ & $  0.33$ & $   0.1092 $ & $   8.7012 $ & $   3.4240$ & $   0.0550 $ & $   0.3258 $ & $  -0.3532 $ & $   0.7324 $ & $  -0.0429 $ & $   4.5242 $ & $ 179.1545$ \\
           & $  100 $ & $  0.26 $ & $  0.31$ & $   0.0844 $ & $   6.8923 $ & $   4.0683$ & $   0.0441 $ & $   0.2985 $ & $  -0.3359 $ & $   0.6027 $ & $   0.0303 $ & $ -14.5100 $ & $ 485.1301$ \\
           & $  250 $ & $  0.25 $ & $  0.31$ & $ 0.0725 $ & $  12.4318 $ & $   3.2776$ & $   0.0557 $ & $   0.2930 $ & $  -0.7418 $ & $   0.3300 $ & $   0.0546 $ & $ -10.3133 $ & $ 823.4443$ \\
           & $ 1000 $ & $  0.24 $ & $  0.31$ & $   0.0562 $ & $  12.9395 $ & $   2.9742$ & $   0.0514 $ & $   0.2925 $ & $  -0.6312 $ & $   0.5611 $ & $   0.0575 $ & $ -10.5586 $ & $ 947.3654$ \\
           & $ 1500 $ & $  0.24 $ & $  0.31$ & $   0.0546 $ & $  13.0970 $ & $   2.9112$ & $   0.0553 $ & $   0.2926 $ & $  -0.7359 $ & $   0.5133 $ & $   0.0573 $ & $ -10.5603 $ & $ 952.6785$ \\
            \hline
           Taus & $  200 $ & $  0.23 $ & $  0.28$ & $   0.0577 $ & $   7.5935 $ & $   3.5566$ & $   0.0341 $ & $   0.2860 $ & $  -0.0818 $ & $   1.4385 $ & $   0.0573 $ & $  -8.8065 $ & $ 935.1002$ \\
           $\chi \chi \rightarrow \tau^+ \tau^-$ & $ 1000 $ & $  0.23 $ & $  0.29$ & $   0.0529 $ & $  12.7237 $ & $   2.9838$ & $   0.0565 $ & $   0.2866 $ & $  -0.8266 $ & $   0.4640 $ & $   0.0562 $ & $ -10.5471 $ & $ 934.1133$ \\
          \hline
   XDM electrons & $   10 $ & $  0.88 $ & $  0.92$ & $   0.2419 $ & $   2.7143 $ & $   4.1521$ & $   0.0908 $ & $   0.4080 $ & $  -0.2529 $ & $   1.1047 $ & $   0.0081 $ & $  -0.9440 $ & $ 149.6370$ \\
   $\chi \chi \rightarrow \phi \phi$ & $  100 $ & $  0.73 $ & $  0.89$ & $   0.2427 $ & $  10.4821 $ & $   3.6656$ & $   0.0792 $ & $   0.3787 $ & $  -0.3787 $ & $   0.6703 $ & $   0.0418 $ & $ -13.7399 $ & $ 296.5718$ \\
   followed by & $  150 $ & $  0.70 $ & $  0.89$ & $   0.2226 $ & $  12.5182 $ & $   3.3474$ & $   0.0686 $ & $   0.3748 $ & $  -0.2138 $ & $   0.7970 $ & $   0.0603 $ & $ -11.9976 $ & $ 292.5551$  \\
   $\phi \rightarrow e^+ e^-$  & $ 1000 $ & $  0.70 $ & $  0.89$ & $   0.1565 $ & $  13.1537 $ & $   2.9202$ & $   0.0727 $ & $   0.3697 $ & $  -0.3598 $ & $   0.6831 $ & $   0.0486 $ & $ -12.7614 $ & $ 675.8390$ \\
   \hline
       XDM muons & $   10 $ & $  0.32 $ & $  0.33$ & $   0.1464 $ & $  23.7835 $ & $   2.7952$ & $   0.0569 $ & $   0.3250 $ & $  -0.4137 $ & $   0.6546 $ & $   0.0370 $ & $  -3.1624 $ & $ 173.1706$ \\
      $\chi \chi \rightarrow \phi \phi$  & $  100 $ & $  0.27 $ & $  0.31$ & $   0.0809 $ & $   2.5357 $ & $   4.7587$ & $   0.0457 $ & $   0.3035 $ & $  -0.3322 $ & $   0.5392 $ & $   0.0179 $ & $ -13.3422 $ & $ 321.8945$ \\
      followed by  & $  400 $ & $  0.25 $ & $  0.31$ & $   0.0741 $ & $  11.3064 $ & $   3.3949$ & $   0.0402 $ & $   0.2937 $ & $  -0.2579 $ & $   0.5965 $ & $   0.0505 $ & $ -10.3800 $ & $ 774.7615$\\
      $\phi \rightarrow \mu^+ \mu^-$   & $ 1000 $ & $  0.25 $ & $  0.31$ & $   0.0617 $ & $  12.5195 $ & $   3.1133$ & $   0.0418 $ & $   0.2925 $ & $  -0.3294 $ & $   0.7487 $ & $   0.0541 $ & $ -10.6936 $ & $ 939.3080$  \\
        & $ 2500 $ & $  0.24 $ & $  0.31$ & $   0.0556 $ & $  13.0389 $ & $   2.9343$ & $   0.0522 $ & $   0.2926 $ & $  -0.6537 $ & $   0.5413 $ & $   0.0566 $ & $ -10.5987 $ & $ 952.4342$\\
       \hline
       XDM taus & $  200 $ & $  0.22 $ & $  0.27$ & $   0.0604 $ & $   6.6206 $ & $   3.6373$ & $   0.0333 $ & $   0.2861 $ & $  -0.0610 $ & $   1.0364 $ & $   0.0548 $ & $  -8.7336 $ & $ 638.6944$ \\ 
      $\chi \chi \rightarrow \phi \phi$, $\phi \rightarrow \tau^+ \tau^-$ & $ 1000 $ & $  0.22 $ & $  0.27$ & $   0.0534 $ & $  11.2208 $ & $   3.1869$ & $   0.0424 $ & $   0.2841 $ & $  -0.4351 $ & $   0.6734 $ & $   0.0542 $ & $ -10.5137 $ & $ 911.3169$ \\ 
     \hline
    XDM pions & $  100 $ & $  0.22 $ & $  0.25$ & $   0.0607 $ & $   1.4685 $ & $   5.0403$ & $   0.0394 $ & $   0.2881 $ & $  -0.2700 $ & $   0.5445 $ & $   0.0137 $ & $ -12.6965 $ & $ 304.5202$ \\
    $\chi \chi \rightarrow \phi \phi$ & $  200 $ & $  0.21 $ & $  0.25$ & $   0.0674 $ & $   6.0060 $ & $   4.1253$ & $   0.0353 $ & $   0.2825 $ & $  -0.1722 $ & $   0.7910 $ & $   0.0323 $ & $ -13.6145 $ & $ 477.7644$  \\
   followed by & $ 1000 $ & $  0.20 $ & $  0.25$ & $   0.0515 $ & $  12.3319 $ & $   3.1745$ & $   0.0382 $ & $   0.2762 $ & $  -0.3601 $ & $   0.6781 $ & $   0.0517 $ & $ -10.8809 $ & $1030.3075$ \\
   $\phi \rightarrow \pi^+ \pi^-$  & $ 1500 $ & $  0.20 $ & $  0.25$ & $   0.0481 $ & $  12.6927 $ & $   3.0715$ & $   0.0428 $ & $   0.2760 $ & $  -0.5297 $ & $   0.5865 $ & $   0.0547 $ & $ -10.7564 $ & $1026.1082$ \\
    & $ 2500 $ & $  0.20 $ & $  0.25$ & $   0.0453 $ & $  12.9871 $ & $   2.9688$ & $   0.0480 $ & $   0.2762 $ & $  -0.6968 $ & $   0.5217 $ & $   0.0566 $ & $ -10.6509 $ & $1025.4334$ \\
   \hline
   			   W bosons & $  200 $ & $  0.29 $ & $  0.35$ & $   0.1013 $ & $  19.1565 $ & $   2.9322$ & $   0.0395 $ & $   0.3076 $ & $  -0.0895 $ & $   1.1093 $ & $   0.0377 $ & $ -13.2287 $ & $ 446.3091$ \\
   				 $\chi \chi \rightarrow W^+ W^-$ & $  300 $ & $  0.29 $ & $  0.35$ & $   0.0906 $ & $  15.7615 $ & $   3.0067$ & $   0.0388 $ & $   0.3053 $ & $  -0.0855 $ & $   1.0554 $ & $   0.0389 $ & $ -13.1812 $ & $ 528.0655$ \\
   			   & $ 1000 $ & $  0.28 $ & $  0.35$ & $   0.0711 $ & $  10.6406 $ & $   3.1935$ & $   0.0415 $ & $   0.3025 $ & $  -0.2181 $ & $   0.8366 $ & $   0.0516 $ & $ -10.0585 $ & $ 782.1619$ \\
   			  \hline
   				 Z bosons & $  200 $ & $  0.28 $ & $  0.34$ & $   0.0998 $ & $  20.7336 $ & $   2.8932$ & $   0.0392 $ & $   0.3043 $ & $  -0.1088 $ & $   1.0375 $ & $   0.0359 $ & $ -13.3227 $ & $ 447.9354$ \\
          $\chi \chi \rightarrow Z Z$ & $ 1000 $ & $  0.27 $ & $  0.33$ & $   0.0689 $ & $  10.6396 $ & $   3.2027$ & $   0.0407 $ & $   0.2988 $ & $  -0.2263 $ & $   0.7934 $ & $   0.0514 $ & $  -9.9893 $ & $ 773.0394$\\
          \hline
       Higgs bosons & $  200 $ & $  0.34 $ & $  0.40$ & $   0.1313 $ & $  24.2160 $ & $   2.8491$ & $   0.0479 $ & $   0.3205 $ & $  -0.2349 $ & $   0.7599 $ & $   0.0297 $ & $ -13.5576 $ & $ 388.8721$ \\
       $\chi \chi \rightarrow h \bar{h}$ & $ 1000 $ & $  0.32 $ & $  0.40$ & $   0.0877 $ & $  10.9585 $ & $   3.1982$ & $   0.0430 $ & $   0.3133 $ & $  -0.1570 $ & $   0.8487 $ & $   0.0490 $ & $  -9.8120 $ & $ 616.1287$  \\
      \hline
           b quarks & $  200 $ & $  0.35 $ & $  0.41$ & $   0.1244 $ & $  20.6286 $ & $   2.8789$ & $   0.0467 $ & $   0.3217 $ & $  -0.1873 $ & $   0.8494 $ & $   0.0345 $ & $ -13.3583 $ & $ 383.5586$ \\
           $\chi \chi \rightarrow b \bar{b}$ & $ 1000 $ & $  0.33 $ & $  0.41$ & $   0.0917 $ & $  11.6611 $ & $   3.1846$ & $   0.0425 $ & $   0.3149 $ & $  -0.1246 $ & $   0.9724 $ & $   0.0467 $ & $  -9.8366 $ & $ 635.3690$ \\
          \hline
 Light quarks & $  200 $ & $  0.34 $ & $  0.40$ & $   0.1129 $ & $  18.5995 $ & $   2.9221$ & $   0.0432 $ & $   0.3174 $ & $  -0.1218 $ & $   0.9244 $ & $   0.0361 $ & $ -13.1747 $ & $ 430.2257$\\
 $\chi \chi \rightarrow u \bar{u}, d \bar{d}$ (50 $\%$ each) & $ 1000 $ & $  0.32 $ & $  0.40$ & $   0.0882 $ & $  12.3648 $ & $   3.1280$ & $   0.0434 $ & $   0.3135 $ & $  -0.1700 $ & $   0.9101 $ & $   0.0490 $ & $  -9.8913 $ & $ 674.5797$ \\
\hline
\end{tabular}
\caption{\label{tab:fitparams}
\emph{Column 3:} Mean values of $f(z)$ averaged from $z=800-1000$. \emph{Column 4:} $f(z=2500)$, where $f$ approaches the asymptotic high-$z$ limit. \emph{Columns 5-7:} Fit parameters for Eq. \ref{eq:lowzfit}; the resulting fit is accurate to within $5 \%$ for $z=10-170$, up to the limits of this calculation. \emph{Columns 8-14:} Fit parameters for Eq. \ref{eq:ffit}; the resulting fit is accurate to within $1 \%$ for $z=170-1470$.
}
\end{table}

At very high redshifts, $z \gg 1000$, $f(z)$ converges to asymptotic values determined solely by the annihilation channel, since in this limit all the energy injected in electrons and photons (as opposed to neutrinos, protons etc) is efficiently deposited. In the (relatively) low-redshift region, $z=10-170$, $f(z)$ can be fitted to within $5 \%$ for all channels considered by a simple three-parameter fit,
\begin{equation} f(z) = a \exp \left( \left( \ln ((1+z)/b) / c \right)^2 \right). \label{eq:lowzfit} \end{equation}
However, at $z \ll 100$, the onset of structure formation may invalidate this calculation. 

Table \ref{tab:fitparams} lists the averaged $f$ employed in Section \ref{sec:cmbconstraints}, the value of $f(z)$ at $z=2500$, and the fit parameters for Eqs \ref{eq:ffit}-\ref{eq:lowzfit}, for the masses and annihilation channels discussed in Subsection \ref{subsec:models}.

\section{Electron Cooling Mechanisms}
\label{app:electrons}

\subsection{Inverse Compton scattering}

The cooling time for ICS is \cite{1970RvMP...42..237B},
\be
\left(\frac{1}{t_\mathrm{cool}}\right) = \frac{- d \ln \gamma}{dt} \sim \frac{4 \sigma_{T} c a_{R} T_{CMB}^{4} 
\gamma}{3 m_{e} c^{2}} \,\,,
\ee
where $T_{CMB} = 2.725 (1+z) {\rm K}$ is the mean CMB temperature at the relevant redshift, $a_{R}$
is the radiation constant, and $\sigma_T = 6.65246 \times 10^{-25}$ cm$^2$ is the Thomson cross section.
Comparing this to the Hubble time, one finds
\be
\frac{t_{H}}{t_\mathrm{cool}} \sim 10^{5} \left( \frac{1+z}{1000} \right)^{5/2}
\frac{1}{\sqrt{\Omega_{M}h^{2}}} \gamma \,\,\,,
\ee
Thus the timescale for ICS is much shorter than the Hubble time.

At the highest energies ($\sim 10$ GeV and higher, depending on the redshift) the Thomson cross section formula is no longer valid. Consequently we must use the general form for the photon spectrum resulting from ICS \cite{1970RvMP...42..237B}. For convenience, all energies are written in units of the electron mass, $m_e c^2 \sim 511$ keV. The doubly differential spectrum is given by,
\begin{equation} \frac{dN}{dE_\gamma d\epsilon dt} = \frac{3}{4} \sigma_T c \frac{1}{\epsilon E_e^2} \left(2 q \log q + (1 + 2 q) (1 - q) + 0.5 (1 - q) (\Gamma q)^2 / (1 + \Gamma q) \right) n(\epsilon) \label{eq:klein-nishina}, \end{equation} 
\[ \Gamma =  4 \epsilon E_e, \]
\[ q = \frac{E_\gamma}{E_e} \frac{1}{\Gamma (1 - E_\gamma/E_e)}. \]
Here $\epsilon$ is the energy (before upscattering) of the soft photon that the electron scatters on: in this case, a CMB photon. $E_e$ is the electron energy, $E_\gamma$ is the energy of the upscattered gamma ray, and $n(\epsilon)$ describes the energy distribution of the soft photons per unit volume (i.e. $n(\epsilon) = dN_\mathrm{CMB}/d\epsilon dV$). This spectrum must then be integrated over the energy distribution of CMB photons to determine the total photon spectrum from ICS of a high-energy electron on the CMB.

\subsection{Ionization, Excitation and Collisional Heating}

We use the fits presented by Arnaud and Rothenflug \cite{1985A&AS...60..425A} for ionization of neutral hydrogen, neutral helium and singly ionized helium,
\begin{equation} \sigma(E_e) = 10^{-14} \mathrm{cm}^2 \frac{1}{u (I / \mathrm{eV})^2} \left\{ A \left( 1 - \frac{1}{u} \right) + B \left(1 - \frac{1}{u} \right)^2 + C \ln u + D \ln u / u \right\} \end{equation}
where $I$ is the ionization potential, $E_e$ is the kinetic energy of the electron and $u = E_e / I$. The coefficients $A$, $B$, $C$, and $D$ are given by,
\begin{itemize}
\item Hydrogen: $A = 22.8$, $B = -12.0$, $C = 1.9$, $D = -22.6$, with $I = 13.6$ eV.
\item Neutral helium: $A = 17.8$, $B = -11.0$, $C = 7.0$, $D = -23.2$, with $I = 24.6$ eV.
\item Singly ionized helium: $A = 14.4$, $B = -5.6$, $C = 1.9$, $D = -13.3$, with $I = 54.4$ eV.
\end{itemize}

For excitation of hydrogen and neutral helium we use the fitting functions given by Stone, Kim and Desclaux \cite{Stone:2002}. For kinetic energies well above threshold, the cross sections have the form,

\begin{equation} \sigma_\mathrm{eH,eHe} = \frac{4 a_0^2 R}{E_\mathrm{kin} + E_\mathrm{bin} + E_\mathrm{exc}} \left(A \ln(E_\mathrm{kin} / R) + B + C R / E_\mathrm{kin} \right). \end{equation}

Here $R \approx 13.6$ eV is the Rydberg energy, $a_0 \approx 0.529 \times 10^{-8}$ cm is the Bohr radius, $E_\mathrm{bin}$ is the binding energy of the electron to be excited, $E_\mathrm{exc}$ is the excitation energy, and $E_\mathrm{kin}$ is the kinetic energy of the incident electron. $A$, $B$ and $C$ are dimensionless parameters with values given by,

\begin{itemize}
\item Hydrogen: $A = 0.5555$, $B=0.2718$, $C=0.0001$,
\item Neutral helium: $A = 0.1771$, $B=-0.0822$, $C=0.0356$.
\end{itemize}

At high energies the result for hydrogen agrees well with that of Shull and van Steenberg \cite{1985ApJ...298..268S},
\be
\sigma_\mathrm{eH} = \frac{2.75 \times 10^{-15} \ln(E/13.6 \mathrm{eV})}{E/\mathrm{eV}} {\rm cm}^2\,\,.
\ee

For excitation of singly ionized helium we follow Fisher et al \cite{PhysRevA.55.329},
\begin{equation} \sigma_{\mathrm{eHe}^+} = \frac{\pi a_0^2}{16 x} \left( \alpha \ln x + \beta \ln x / x + \gamma + \delta / x + \eta / x^2 \right), \end{equation}
\[ x = E_\mathrm{kin}/E_\mathrm{exc}, \]
\[ \alpha = 3.22, \, \beta = 0.357, \, \gamma = 0.00157, \, \delta = 1.59, \, \eta = 0.764. \]

Collisional losses become important
at lower energies, $100\eV < E < 1\keV$. The cross section for collisional heating is given by \cite{1985ApJ...298..268S},
\begin{equation} \sigma_{ee} = (7.82 \times 10^{-11} ) (0.05/f) (\ln \Lambda ) (E_e / \mathrm{eV})^{-2} \mathrm{cm}^2, \end{equation}
where $f \equiv \Delta E / E = 0.05$ is chosen to simulate the discrete nature of Coulomb collisions. We take the Coulomb logarithm $\ln \Lambda \sim 10$.

\section{Photon Cooling Mechanisms}
\label{app:photons}

\subsection{Pair production on the CMB}

The doubly differential electron spectrum for pair production by a gamma ray (energy $E_\gamma$) encountering a soft photon (energy $\epsilon$) is given by \cite{1983Ap.....19..187A, Ferrigno:2004am},
\begin{equation} \frac{dN}{dE_e d\epsilon dt} = \sigma_T c \frac{3}{64} \frac{1}{\epsilon^2 E_\gamma^3} \left(\frac{4 A^2 \ln (4 \epsilon E_e (A - E_e)/A)}{E_e (A-E_e)} - 8 \epsilon A + \frac{2 (2 \epsilon A -1) A^2}{E_e (A-E_e)} - \left(1-\frac{1}{\epsilon A}\right) \frac{A^4}{E_e^2 (A-E_e)^2} \right) n(\epsilon) \end{equation}
\[ \frac{A}{2} \left(1 - \sqrt{1-\frac{1}{E_\gamma \epsilon}} \right) < E_e < \frac{A}{2} \left(1 + \sqrt{1-\frac{1}{E_\gamma \epsilon}} \right) \]
Here $A = E_\gamma + \epsilon$. As in Eq. \ref{eq:klein-nishina}, all energies are in units of the electron mass, and $n(\epsilon)$ describes the energy distribution of the soft photons per unit volume. This expression must be integrated over the CMB energy distribution to obtain the spectrum of electrons produced by pair production from a high energy gamma ray on the CMB. The positron spectrum is identical.

\subsection{Pair production on H and He}

We employ the high energy (Born approximation) cross sections for pair production on ionized H, free electrons, and singly ionized He, as described by Motz, Olsen and Koch \cite{1969RvMP...41..581M}. As previously, we write all energies in units of the electron mass. In the high energy limit (complete screening), He$^+$ and H$^+$ can both be regarded as singly charged point charges, so share a single cross section for pair production,
\begin{equation} \sigma = \alpha r_0^2 \left( \frac{28}{9} \ln (2 E_\gamma) - \frac{218}{27} \right), \end{equation}
where $\alpha$ is the fine structure constant and $r_0 = 2.8179 \times 10^{-13}$ cm is the classical electron radius. If this equation is also used to describe pair production in the field of free electrons, then in the limit of full ionization (including double ionization of helium), this expression agrees with that given in \cite{1989ApJ...344..551Z} for a fully ionized medium. However, we slightly modify this formula as suggested by Joseph and Rohrlich \cite{RevModPhys.30.354} for pair production off electrons,
\begin{equation} \sigma = \alpha r_0^2 \left( \frac{28}{9} \ln (2 E_\gamma) - \frac{100}{9} \right). \end{equation}

For pair production on neutral hydrogen and helium, we employ the cross sections given in \cite{1989ApJ...344..551Z},
\begin{equation} \sigma_\mathrm{H} = 5.4 \alpha r_0^2 \ln \frac{513 E_\gamma}{E_\gamma + 825}, \end{equation}
\begin{equation} \sigma_\mathrm{He} = 8.76 \alpha r_0^2 \ln \frac{513 E_\gamma}{E_\gamma + 825}. \end{equation}

In all cases we use the high energy Born approximation form for the spectrum of produced pairs \cite{1969RvMP...41..581M}.

\subsection{Photon-photon scattering}

Photon-photon scattering occurs when a gamma ray upscatters a CMB photon. It can be regarded as a photon ``splitting'' process, with each photon in the final state carrying away $\sim 1/2$ the energy of the original gamma ray. We use the rate and spectrum given in \cite{1990ApJ...349..415S}.

The total rate for scattering of a gamma ray of energy $E_\gamma$ on the CMB (already integrated over the CMB spectrum) is given by,
\begin{equation} R = 1.83 \times 10^{-27} h_{50}^{-1} T_{2.7}^6 H_0 (1+z)^6 E_\gamma^3, \end{equation}
where $T_{2.7}$ is the present CMB temperature divided by $2.7$ K, $E_\gamma$ is measured in units of the electron mass, and $h_{50} = H_0 / (50 \mathrm{km /s / Mpc}) \sim 1.4$. The spectrum of resulting gamma rays is given by,
\begin{equation} \frac{dN}{dE dt} = R (20/7) \frac{1}{E_\gamma} \left(1 - \frac{E}{E_\gamma} + \left(\frac{E}{E_\gamma}\right)^2 \right)^2. \end{equation}

\subsection{Compton scattering}

Following \cite{Chen:2003gz}, we treat all electrons as free for Compton processes. The differential cross section for Compton scattering is just the usual Klein-Nishina cross section \cite{1954qtr..book.....H}, given by,

\begin{equation} \frac{d\sigma}{d E_\gamma'} = \pi \alpha^2 r_c^2 \frac{1}{E_\gamma^2} \left(\frac{E_\gamma'}{E_\gamma} + \frac{E_\gamma}{E_\gamma'} -1 +  \left(1 - \left(\frac{1}{E_\gamma'} - \frac{1}{E_\gamma} \right) \right)^2 \right),\end{equation}
\begin{equation} \frac{E_\gamma}{1 + 2 E_\gamma} < E_\gamma' < E_\gamma, \end{equation}
where $r_c = 3.862 \times 10^{-11}$ cm is the Compton radius of the electron, $E_\gamma$ is the initial energy of the photon, $E_\gamma'$ is its final energy, and all energies are in units of the electron mass.

\subsection{Photoionization}

We include photoionization on neutral hydrogen, neutral helium and singly ionized helium. It is crucial to include photoionization on helium, as otherwise photoionization appears to turn off at redshifts greater than $z \sim 1000$ (i.e. hydrogen recombination), which does not reflect the true situation.

The photoionization cross sections for hydrogen and singly ionized helium are known analytically, and are given by \cite{1989ApJ...344..551Z},
\begin{eqnarray} \eta & = & \frac{1}{\sqrt{E/E_\mathrm{thres} - 1}} \nonumber \\
\sigma & = & \frac{2^9 \pi^2 r_0^2}{3 \alpha^3} (E_\mathrm{thres}/E)^4 \frac{\exp (-4 \eta \arctan(1/\eta))}{1 - \exp(-2 \pi \eta)}, \end{eqnarray}
where $r_0 =  2.818 \times 10^{-13}$ cm is the classical electron radius, and $E_\mathrm{thres} = 13.6$ eV for H, $54.4$ eV for He$^+$.

Svensson and Zdziarski \cite{1989ApJ...344..551Z} fitted the photoionization cross section for a mixture of $75 \%$ neutral hydrogen and $25 \%$ neutral helium (by mass) with a broken power law. We subtract the hydrogen contribution and use the result as the photoionization cross section for neutral helium,
\begin{equation} \sigma(\mathrm{He}) = -12 \sigma (\mathrm{H}) + 5.1 \times 10^{-20} \mathrm{cm}^2 \left\{ \begin{array}{cc}  \left(\frac{E}{250 \mathrm{eV}}\right)^{-2.65}, & \quad 50 \mathrm{eV} \, < \, E \, < 250 \mathrm{eV}\\  \left(\frac{E}{250 \mathrm{eV}}\right)^{-3.30}, & \quad E \, > 250 \mathrm{eV}. \end{array} \right. \end{equation}

\bibliography{epsilon}

\end{document}